\documentclass[aps,preprint]{revtex4}%
\usepackage{amsmath}
\usepackage{amsfonts}
\usepackage{amssymb}
\usepackage{graphicx}
\usepackage{mathtools}
\usepackage{graphics}
\usepackage{epsfig}
\usepackage[section]{placeins}
\usepackage{float}
\usepackage{multirow}
\usepackage{bibentry}
\usepackage{braket}
\usepackage[font=footnotesize,labelfont=bf]{caption}
\usepackage{pdfpages}
\usepackage{slashbox}%
\providecommand{\U}[1]{\protect\rule{.1in}{.1in}}
\nobibliography*

\begin{document}
\preprint{ }
\title{Causality bounds for neutron-proton scattering}
\author{Serdar Elhatisari and Dean Lee}
\affiliation{Department of Physics, North Carolina State University, Raleigh, NC 27695,
USA\linebreak}

\begin{abstract}
We consider the constraints of causality and unitarity for the low-energy
interactions of protons and neutrons. We derive a general theorem that
non-vanishing partial-wave mixing cannot be reproduced with zero-range
interactions without violating causality or unitarity. \ We define and
calculate interaction length scales which we call the causal range and the
Cauchy-Schwarz range for all spin channels up to $J=3$. For some channels we
find that these length scales are as large as $5$ fm. \ We investigate the
origin of these large lengths and discuss their significance for the choice of
momentum cutoff scales in effective field theory and universality in many-body
Fermi systems.

\end{abstract}
\maketitle
\preprint{ }

\section{Introduction}

\label{sec:introduction} Chiral effective field theory describes the
low-energy interactions of protons and neutrons. \ If one neglects
electromagnetic effects, the long range behavior of the nuclear interactions
is determined by pion exchange processes. \ See
Ref.~\cite{vanKolck:1999mw,Bedaque:2002mn,Epelbaum:2005pn,Epelbaum:2008ga} for
reviews on chiral effective field theory. \ But there are also systems of
interest where momenta smaller than the pion mass are relevant. \ In such
cases it is more economical to use pionless effective field theory with only
local contact interactions involving the nucleons. \ The pionless formulation
is theoretically elegant since the theory at leading order is renormalizable
and the momentum cutoff scale can be arbitrarily large
\cite{vanKolck:1998bw,Bedaque:1998kg,Bedaque:1998km,Bedaque:1999ve,Chen:1999tn,Platter:2004zs,
Hammer:2006ct}. \ This allows an elegant connection with the universal
low-energy physics of fermions at large scattering length and other systems
such as ultracold atoms \cite{Braaten:2004a,Giorgini:2007a}.

For local contact interactions the range of the interactions are set by the
momentum cutoff scale for the effective theory. \ There are rigorous
constraints for strictly finite-range interactions set by causality and
unitarity. \ Some violations of unitarity can relax these constraints if one
works at finite order in perturbation theory or includes unphysical
propagating modes with negative norm. \ However at some point one must
accurately reproduce the underlying unitary quantum system by going to
sufficiently high order in perturbation theory or decoupling the effects of
propagating unphysical modes.

The time evolution of any quantum mechanical system obeys causality and
unitarity. \ Causality requires that the cause of an event must occur before
any resulting consequences are produced, and unitarity requires that the sum
of all outcome probabilities equals one. \ In the case of non-relativistic
scattering, these constraints mean that the outgoing wave may depart only
after the incoming wave reaches the scattering object and must preserve the
normalization of the incoming wave. \ In this paper we discuss the constraints
of causality and unitarity for finite range interactions. \ Specifically we
consider neutron-proton scattering in all spin channels up to $J=3$.

The constraints of causality for finite-range interactions were first
investigated by Wigner \cite{Wigner1954}. \ The time delay between an incoming
wave packet and the scattered outgoing wave packet is equal to the energy
derivative of the elastic phase shift, $\Delta t=2\hbar\,d\delta/dE$. \ If
$d\delta/dE$ is negative, the outgoing wave is produced earlier than that for
the non-interacting system. \ However the incoming wave must first arrive in
the interacting region before the outgoing wave can be produced. \ For each
partial wave, $L$, this puts an upper bound on the effective range parameter,
$r_{L}$, in the effective range expansion,%
\begin{equation}
k^{2L+1}\cot\delta_{L}(k)=-\frac{1}{a_{L}}+\frac{1}{2}r_{L}k^{2}+O(k^{4}).
\end{equation}
\ Phillips and Cohen \cite{Phillips1997} derived the causality bound for the
$S$-wave effective range parameter for finite-range interactions in three
dimensions. \ Further work on causality bounds and universal relations at low
energies have been carried out by Ruiz Arriola and collaborators.
\ Constraints on nucleon-nucleon scattering and the chiral two-pion exchange
potential was considered in Ref.~\cite{PavonValderrama:2005wv}. \ Correlations
between the scattering length and effective range were discussed for one boson
exchange potentials \cite{Cordon:2009pj} and van der Waals potentials
\cite{Cordon:2010a}. \

The causality bounds for arbitrary dimension $d$ and arbitrary angular
momentum $L$ were derived in Ref.~\cite{Hammer-Lee9,HammerDean27}. \ Let $R$
be the range of the interaction. \ For the case $d=3$, it was found that the
effective range parameter must satisfy the upper bound%
\begin{align}
r_{L}  &  \leq b_{L}(r)=-\frac{2\Gamma(L-\frac{1}{2})\Gamma(L+\frac{1}{2}%
)}{\pi}\left(  \frac{r}{2}\right)  ^{-2L+1}\nonumber\\
&  \qquad-\frac{4}{L+\frac{1}{2}}\frac{1}{a_{L}}\left(  \frac{r}{2}\right)
^{2}+\frac{2\pi}{\Gamma(L+\frac{3}{2})\Gamma(L+\frac{5}{2})}\frac{1}{a_{L}%
^{2}}\left(  \frac{r}{2}\right)  ^{2L+3}, \label{bL}%
\end{align}
for any $r\geq R$. \ This inequality can be used to determine a length scale,
$R^{b}$, which we call the causal range,%
\begin{equation}
r_{L}=b_{L}(R^{b}).
\end{equation}
The physical meaning of $R^{b}$ is that any set of interactions with strictly
finite range that reproduces the physical scattering data must have a range
greater than or equal to $R^{b}$.

In this paper we extend the causality bound and causal range and to the case
of two coupled partial-wave channels. \ For applications to nucleon-nucleon
scattering the relevant coupled channels are $^{3}S_{1}$-$^{3}D_{1}$,
$^{3}P_{2}$-$^{3}F_{2}$, $^{3}D_{3}$-$^{3}G_{3},$ etc. \ As we will show,
there is some modification of the effective range bound in Eq.~(\ref{bL}) due
to mixing. \ For total spin $J$ we show that the lower partial-wave channel
$L=J-1$ satisfies the new causality bound,%
\begin{equation}
r_{J-1}\leq b_{J-1}(r)-2q_{0}^{2}\frac{\Gamma(J+\frac{1}{2})\Gamma(J+\frac
{3}{2})}{\pi}\left(  \frac{r}{2}\right)  ^{-2J-1}, \label{preview_r_J-1}%
\end{equation}
where $q_{0}$ is the first term in the expansion of the mixing angle
$\varepsilon_{J}$ in the Blatt-Biedenharn eigenphase convention
\cite{Blatt-B.I},%
\begin{equation}
\tan\varepsilon_{J}(k)=q_{0}k^{2}+q_{1}k^{4}+O(k^{6}). \label{taneps}%
\end{equation}
We note that the last term in Eq.~(\ref{preview_r_J-1}) is negative
semi-definite and diverges as $r\rightarrow0$. \ From this observation we make
the general statement that non-vanishing partial-wave mixing is inconsistent
with zero-range interactions. \ We will explore in detail the consequences of
this result as it applies to nuclear effective field theory.

We also derive a new causality bound associated with the mixing angle itself.
\ Using the Cauchy-Schwarz inequality we derive a bound for the parameter
$q_{1}$ in the expansion Eq.~(\ref{taneps}). \ This leads to another minimum
interaction length scale, which we call the Cauchy-Schwarz range,
$R^{\text{C-S}}$. We use the new causality bounds to determine the minimum
causal and Cauchy-Schwarz ranges for each $^{2S+1}L_{J}$ channel in
neutron-proton scattering up to $J=3$. \ Since the long range behavior of the
nuclear interactions is determined by pion exchange processes, one expects
$R^{b}\sim R^{\text{C-S}}\sim m_{\pi}^{-1}=1.5$ fm. \ However in some higher
partial-wave channels we find that these length scales are as large as $5$~fm.
\ We show these large ranges are generated by the one-pion exchange tail.
\ Using a potential model we show that the causal range and Cauchy-Schwarz
range are both significantly reduced when the one-pion exchange tail is
chopped off at distances beyond $5$~fm. \ We discuss the impact of this
finding on the choice of momentum cutoff scales in effective field theory.

In the limit of isospin symmetry our analysis of the isospin triplet channels
can also be applied to neutron-neutron scattering and therefore has relevance
to dilute neutron matter. \ The physics of dilute neutron matter is important
for describing the crust of neutron stars as well as connections to the
universal physics of fermions near the unitarity limit. \ Our analysis of the
causality and unitarity bounds show that there are constraints on the
universal character of neutron-neutron interactions in channels with
partial-wave mixing as well as higher uncoupled partial-wave channels. \ In
other words some low-energy phenomenology cannot be cleanly separated from
microscopic details such the range of the interaction. \ Reviews of the theory
of ultracold Fermi gases close to the unitarity limit and their numerical
simulations are given in Ref.~\cite{Giorgini:2007a,Lee:2008fa}. A general
overview of universality at large scattering length can be found in
Ref.~\cite{Braaten:2004a}. \ See Ref.~\cite{Koehler:2006A, Regal:2006thesis}
for reviews of recent cold atom experiments at unitarity.

\section{Uncoupled Channels}

We analyze in this section the channels with only one partial wave, $L$. \ We
summarize the results obtained in Ref. \cite{HammerDean27}. \ For simplicity,
we will assume throughout the calculations that the interaction has finite
range $R$, and we use units where $\hbar=1$. \ For the two-body system the
total wave function in the relative coordinate is%
\begin{equation}
\Psi^{(k)}(\overset{\rightharpoonup}{r})=R_{L}^{(k)}(r)\text{Y}_{L,M_{L}%
}(\theta,\phi). \label{eqn:uncoup:TotalWave}%
\end{equation}
$R_{L}^{(k)}(r)$ is the radial part, and the $\text{Y}_{L,M_{L}}(\theta,\phi)$
are the spherical harmonics. The radial wave function $R_{L}^{(k)}(r)$
satisfies the radial Schr\"{o}dinger equation,
\begin{equation}
\left[  -\frac{1}{r^{2}}\frac{d}{dr}\left(  r^{2}\frac{d}{dr}\right)
+\frac{L(L+1)}{r^{2}}-k^{2}\right]  R_{L}^{(k)}(r)+2\mu\int_{0}^{R}%
W(r,r^{\prime})U_{L}^{(k)}(r^{\prime})dr^{\prime}=0.
\label{eqn:uncoup:radialeqR}%
\end{equation}
Here $\mu$ is the reduced mass. \ We write $W(r,r^{\prime})$ for the non-local
interaction potential as a real symmetric integral operator. \ Since the
interaction has finite range $R$, we require that $W(r,r^{\prime})=0$ for
$r>R$ or $r^{\prime}>R$. $U_{L}^{(k)}(r)$ is the rescaled the radial
function,
\begin{equation}
R_{L}^{(k)}(r)=\frac{U_{L}^{(k)}(r)}{kr}. \label{eqn:rescaledU}%
\end{equation}
The effective range expansion for channels with a single partial wave is,
\begin{equation}
k^{2L+1}\cot\delta_{L}(k)=\frac{-1}{a_{L}}+\frac{1}{2}r_{L}k^{2}%
+\mathcal{O}\left(  k^{4}\right)  . \label{eqn:oneChanneleffecranexp}%
\end{equation}

It is shown in Ref. \cite{HammerDean27} that for any $r>R$ the effective range
satisfies
\begin{equation}
r_{L}=b_{L}(r)-2\int_{0}^{r}\left[  U_{L}^{(0)}(r^{\prime})\right]  {}%
^{2}dr^{\prime}, \label{eqn:uncoup:effectrang1}%
\end{equation}
where $b_{L}(r)$ is
\begin{align}
b_{L}(r)=  &  \frac{1}{a_{L}^{2}}\frac{2\pi}{\Gamma\left(  L+\frac{3}%
{2}\right)  \Gamma\left(  L+\frac{5}{2}\right)  }\left(  \frac{r}{2}\right)
^{2L+3}\nonumber\\
&  -\frac{1}{a_{L}}\frac{4}{L+\frac{1}{2}}\left(  \frac{r}{2}\right)
^{2}-\frac{2\Gamma\left(  L-\frac{1}{2}\right)  \Gamma\left(  L+\frac{1}%
{2}\right)  }{\pi}\left(  \frac{r}{2}\right)  ^{-2L+1}.
\label{eqn:uncoup:generalb}%
\end{align}
Since the wave function is real and the integral term in
Eq.~(\ref{eqn:uncoup:effectrang1}) is positive semi-definite, this equation
puts an upper bound on the effective range, $r_{L}\leq b_{L}(r)$.

\section{Coupled Channels}

\label{sec:coupledChannel}

In this section we derive the general wave functions for spin-triplet
scattering with mixing between orbital angular momentum $L=J-1$ and $L=J+1$.
The coupled-channel wave functions satisfy the following coupled radial
Schr\"{o}dinger equations,%
\begin{align}
&  \left[  -\frac{d^{2}}{dr^{2}}-k^{2}+\frac{J(J-1)}{r^{2}}\right]
U_{J-1}^{(k)}(r)\nonumber\\
&  +2\mu\int_{0}^{R}[W_{11}(r,r^{\prime})U_{J-1}^{(k)}(r^{\prime}%
)+W_{12}(r,r^{\prime})V_{J+1}^{(k)}(r^{\prime})]\,dr=0,
\label{eqn:couplediff1}%
\end{align}%
\begin{align}
&  \left[  -\frac{d^{2}}{dr^{2}}-k^{2}+\frac{(J+1)(J+2)}{r^{2}}\right]
V_{J+1}^{(k)}(r)\nonumber\\
&  +2\mu\int_{0}^{R}[W_{21}(r,r^{\prime})U_{J-1}^{(k)}(r^{\prime}%
)+W_{22}(r,r^{\prime})V_{J+1}^{(k)}(r^{\prime})]\,dr^{\prime}=0.
\label{eqn:couplediff2}%
\end{align}
Here the non-local interaction potentials are represented by a real symmetric
$2\times2$ matrix $W(r,r^{\prime})$,
\begin{equation}
W(r,r^{\prime})=\left(
\begin{array}
[c]{cc}%
W_{11}(r,r^{\prime}) & W_{12}(r,r^{\prime})\\
W_{12}(r,r^{\prime}) & W_{22}(r,r^{\prime})
\end{array}
\right)  . \label{eqn:PotentialMatrix}%
\end{equation}
In Eq.~(\ref{eqn:couplediff1})-(\ref{eqn:couplediff2}) the $U_{J-1}^{(k)}(r)$
corresponds with the spin-triplet $L=J-1$ channel and the $V_{J+1}^{(k)}(r)$
is for the spin-triplet $L=J+1$. These wave functions are the rescaled form of
the radial wave functions. \ In the non-interacting region $r\geq R$ the
coupled radial Schr\"{o}dinger equations reduce to the free radial
Schr\"{o}dinger equations
\begin{equation}
\left(  -\frac{d^{2}}{dr^{2}}-k^{2}+\frac{J(J-1)}{r^{2}}\right)  U_{J-1}%
^{(k)}(r)=0, \label{eqn:freediff1}%
\end{equation}%
\begin{equation}
\left(  -\frac{d^{2}}{dr^{2}}-k^{2}+\frac{(J+1)(J+2)}{r^{2}}\right)
V_{J+1}^{(k)}(r)=0. \label{eqn:freediff2}%
\end{equation}
The solutions of these differential equations are the Riccati-Bessel
functions,
\begin{equation}
U_{J-1}^{(k)}(r)=A_{1}S_{J-1}(kr)+B_{1}C_{J-1}(kr),
\label{eqn:uwave1}%
\end{equation}%
\begin{equation}
V_{J+1}^{(k)}(r)=A_{2}S_{J+1}(kr)+B_{2}C_{J+1}(kr).
\label{eqn:vwave1}%
\end{equation}
where $A_{1,2}$ and $B_{1,2}$ are amplitudes associated with incoming and
outgoing waves, respectively. More details regarding the Riccati-Bessel functions are given in Appendix
\ref{append:RiccatiBessel}. \ The relation between incoming and outgoing wave
amplitudes is
\begin{equation}
B=\hat{\mathbf{K}}A, \label{eqn:matr1}%
\end{equation}
$\hat{\mathbf{K}}^{-1}$ is the reaction matrix and is defined in terms of the unitary scattering matrix $S$
\begin{equation}
\hat{\mathbf{K}}=i(1-S)(1+S)^{-1}.
\end{equation}
Therefore, the Eq.~(\ref{eqn:matr1}) is written as
\begin{equation}
\tilde{B}=S \tilde{A}, \label{eqn:matr2}%
\end{equation}
where $\tilde{A}_{1,2}$ and $\tilde{B}_{1,2}$ are rescaled amplitudes associated with incoming and
outgoing waves. \ For two coupled channels the
$2\times2$ scattering matrix can also be made symmetric. \ It is possible to
write several different $2\times2$ $S$-matrices which satisfy the unitarity
and symmetry properties. In the literature, there are two conventionally used
$2\times2$ $S$-matrices \cite{Stapp1956, Blatt-B.I}. In this study we adopt
the \textquotedblleft eigenphase" parameterizations of Blatt and Biedernharn
\cite{Blatt-B.I}, and the relations between the eigenphase and nuclear bar
\cite{Stapp1956} parameterizations are shown in Appendix \ref{append:B}.

The $S$-matrix can be diagonalized by an orthogonal matrix $U$
\begin{equation}
S_{d}=USU^{-1}=\left(
\begin{array}
[c]{cc}%
e^{2i\delta_{\alpha}} & 0\\
0 & e^{2i\delta_{\beta}}%
\end{array}
\right)  ,\label{eqn:sdiag}%
\end{equation}
that contains one real parameter $\varepsilon,$%
\begin{equation}
U=\left(
\begin{array}
[c]{cc}%
\cos\varepsilon & \sin\varepsilon\\
-\sin\varepsilon & \cos\varepsilon
\end{array}
\right)  .\label{eqn:unitmatr}%
\end{equation}
$\delta_{\alpha}(k)$ and $\delta_{\beta}(k)$ are the two phase shifts, and
$\varepsilon(k)$ is the mixing angle. The $S$-matrix explicitly is%
\begin{equation}
S=\left(
\begin{array}
[c]{cc}%
e^{2i\delta_{\alpha}}\cos^{2}\varepsilon+e^{2i\delta_{\beta}}\sin
^{2}\varepsilon & \cos\varepsilon\sin\varepsilon\left(  e^{2i\delta_{\alpha}%
}-e^{2i\delta_{\beta}}\right)  \\
\cos\varepsilon\sin\varepsilon\left(  e^{2i\delta_{\alpha}}-e^{2i\delta
_{\beta}}\right)   & e^{2i\delta_{\alpha}}\sin^{2}\varepsilon+e^{2i\delta
_{\beta}}\cos^{2}\varepsilon
\end{array}
\right)  .\label{eqn:scattmat}%
\end{equation}

The eigenvalue equation $S\Ket{X}=\lambda\Ket{X}$ results in eigenvalues
$\lambda_{1}=e^{2i\delta_{\alpha}}$ and $\lambda_{2}=e^{2i\delta_{\beta}}$,
with corresponding eigenstates,
\begin{equation}
\Ket{X_1}=\left(
\begin{array}
[c]{c}%
\cos\varepsilon\\
\sin\varepsilon
\end{array}
\right)  \ \ \text{and}\ \ \Ket{X_2}=\left(
\begin{array}
[c]{c}%
-\sin\varepsilon\\
\cos\varepsilon
\end{array}
\right)  ,
\end{equation}
which satisfy the orthogonality condition
\begin{equation}
\Braket{X_1 | X_2}=0. \label{eqn:orthg}%
\end{equation}
We can write Eq.~(\ref{eqn:matr2}) as
\begin{equation}
\left(
\begin{array}
[c]{cc}%
\tilde{B}_{1\alpha} & \tilde{B}_{1\beta}\\
\tilde{B}_{2\alpha} & \tilde{B}_{2\beta}%
\end{array}
\right)  =\left(
\begin{array}
[c]{cc}%
S_{11} & S_{12}\\
S_{12} & S_{22}%
\end{array}
\right)  \left(
\begin{array}
[c]{cc}%
\tilde{A}_{1\alpha} & \tilde{A}_{1\beta}\\
\tilde{A}_{2\alpha} & \tilde{A}_{2\beta}%
\end{array}
\right)  , \label{eqn:matr3}%
\end{equation}
where the matrices $\tilde{A}$ and $\tilde{B}$ are
\begin{equation}
\tilde{A}=\left(
\begin{array}
[c]{cc}%
e^{-i\delta_{\alpha}}\cos\varepsilon & -e^{-i\delta_{\beta}}\sin\varepsilon\\
e^{-i\delta_{\alpha}}\sin\varepsilon & e^{-i\delta_{\beta}}\cos\varepsilon
\end{array}
\right)  , \label{eqn:Amatrix}%
\end{equation}%
\begin{equation}
\tilde{B}=\left(
\begin{array}
[c]{cc}%
e^{i\delta_{\alpha}}\cos\varepsilon & -e^{i\delta_{\beta}}\sin\varepsilon\\
e^{i\delta_{\alpha}}\sin\varepsilon & e^{i\delta_{\beta}}\cos\varepsilon
\end{array}
\right)  .\hspace{0.4cm} \label{eqn:Bmatrix}%
\end{equation}

We now define some additional notation. \ We write all $\alpha$-state
phaseshifts $\delta_{\alpha}(k)$ as $\delta_{J-1}(k)$ and all $\beta$-state
phaseshifts $\delta_{\beta}(k)$ as $\delta_{J+1}(k)$. \ The notation is
appropriate since in the $k\rightarrow0$ limit the $\alpha$-state is purely
$L=J-1$ and the $\beta$-state is purely $L=J+1$. \ We also drop the
superscript $k$ in the wave functions. \ We choose the normalization of the
wave function to be well-behaved in the zero-energy limit. \ Using the
relations%
\begin{equation}
S_{J\pm1}(kr)\underset{\text{as}\ k\rightarrow0}{\longrightarrow}\sqrt{\pi
}(kr)^{J\pm1+1}\frac{2^{-J\mp1-1}}{\Gamma(J\pm1+3/2)},
\label{eqn:firstricbes2}%
\end{equation}%
\begin{equation}
C_{J\pm1}(kr)\underset{\text{as}\ k\rightarrow0}{\longrightarrow}%
\frac{(kr)^{-J\mp1}}{\sqrt{\pi}}2^{J\pm1}\Gamma(J\pm1+1/2),
\label{eqn:secondricbes2}%
\end{equation}
and removing an overall phase factor, we get wave functions of the form%
\begin{align}
&  U_{\alpha}(r)=\cos\varepsilon_{J}(k)\ k^{J-1}[\cot\delta_{J-1}%
(k)S_{J-1}(kr)+C_{J-1}(kr)],\label{eqn:ualpwave3}\\
&  V_{\alpha}(r)=\sin\varepsilon_{J}(k)\ k^{J-1}[\cot\delta_{J-1}%
(k)S_{J+1}(kr)+C_{J+1}(kr)],\\
&  U_{\beta}(r)=-\sin\varepsilon_{J}(k)\ k^{J+1}[\cot\delta_{J+1}%
(k)S_{J-1}(kr)+C_{J-1}(kr)],\\
&  V_{\beta}(r)=\cos\varepsilon_{J}(k)\ k^{J+1}[\cot\delta_{J+1}%
(k)S_{J+1}(kr)+C_{J+1}(kr)]. \label{eqn:vbetwave3}%
\end{align}

For later convenience we define
\begin{align}
s_{J\pm1}(k,r)  &  =k^{-J\mp1-1}S_{J\pm1}(kr),\label{eqn:expfuncS}\\
c_{J\pm1}(k,r)  &  =k^{J\pm1}C_{J\pm1}(kr).\hspace{0.4cm} \label{eqn:expfuncC}%
\end{align}
Eq.~(\ref{eqn:expfuncS})-(\ref{eqn:expfuncC}) with Eq.~(\ref{eqn:firstricbes}%
)-(\ref{eqn:secondricbes}) indicate that $s_{J\pm1}(k,r)$ and $c_{J\pm1}(k,r)$
are analytic functions of $k^{2}$, and they can be written as
\begin{align}
s_{J\pm1}(k,r)  &  =s_{0,J\pm1}(r)+k^{2}s_{2,J\pm1}(r)+\mathcal{O}\left(
k^{4}\right)  ,\label{eqn:expfuncS2}\\
c_{J\pm1}(k,r)  &  =c_{0,J\pm1}(r)+k^{2}c_{2,J\pm1}(r)+\mathcal{O}\left(
k^{4}\right)  . \label{eqn:expfuncC2}%
\end{align}
The explicit form for the terms in these expansions are given in Appendix
\ref{append:RiccatiBessel}. Therefore Eq.~(\ref{eqn:ualpwave3}%
)-(\ref{eqn:vbetwave3}) become
\begin{align}
&  U_{\alpha}(r)=\cos\varepsilon_{J}(k)[k^{2J-1}\cot\delta_{J-1}%
(k)s_{J-1}(k,r)+c_{J-1}(k,r)],\label{eqn:ualpwave4}\\
&  V_{\alpha}(r)=\sin\varepsilon_{J}(k)[k^{2J+1}\cot\delta_{J-1}%
(k)s_{J+1}(k,r)+k^{-2}c_{J+1}(k,r)], \label{eqn:valphwave4}%
\end{align}%
\begin{align}
&  U_{\beta}(r)=-\sin\varepsilon_{J}(k)[k^{2J+1}\cot\delta_{J+1}%
(k)s_{J-1}(k,r)+k^{2}c_{J-1}(k,r)],\label{eqn:ubetwave4}\\
&  V_{\beta}(r)=\cos\varepsilon_{J}(k)[k^{2J+3}\cot\delta_{J+1}(k)s_{J+1}%
(k,r)+c_{J+1}(k,r)]. \label{eqn:vbetwave4}%
\end{align}

Analogous to the effective range expansion defined in
Eq.~(\ref{eqn:oneChanneleffecranexp}), the two-channel effective range
expansion has the following form,
\begin{equation}
k^{L_{ij}+\frac{1}{2}}\hat{\mathbf{K}}^{-1}k^{L_{ij}+\frac{1}{2}}=-\frac
{1}{\mathbf{a}_{L_{ij}}}+\frac{1}{2}\mathbf{r}_{L_{ij}}k^{2}+\mathcal{O}%
(k^{4}),
\end{equation}
where $\mathbf{a}_{L_{ij}}$ is the scattering length matrix, $\mathbf{r}_{\ell_{ij}%
}$ is the effective range matrix, and $k^{L_{ij}}$ is the diagonal momentum
matrix $\text{diag}(k^{J-1},k^{J+1})$. The two-channel effective range
expansion in the Blatt and Biedernharn parameterization is%
\begin{equation}
k^{L_{ij}+\frac{1}{2}}U\hat{\mathbf{K}}^{-1}U^{-1}k^{L_{ij}+\frac{1}{2}%
}=-\frac{1}{a_{L_{ij}}}+\frac{1}{2}r_{L_{ij}}k^{2}+\mathcal{O}(k^{4})
\end{equation}
where $a_{L_{ij}}=\text{diag}(a_{J-1},a_{J+1})$, and $r_{L_{ij}}%
=\text{diag}(r_{J-1},r_{J+1})$. In addition we get an analytic expansion for
the tangent mixing angle \cite{Blatt-B.II}
\begin{equation}
\tan\varepsilon_{J}(k)=q_{0}k^{2}+q_{1}k^{4}+\mathcal{O}\left(  k^{6}\right)
\label{eqn:mixingangleexpans}%
\end{equation}
with mixing parameters $q_{0}$ and $q_{1}$. \ Now using
Eq.~(\ref{eqn:mixingangleexpans}) we obtain the following final forms of wave
functions for $r\geq R$,%
\begin{align}
U_{\alpha}(r)  &  =\frac{-1}{a_{J-1}}s_{0,J-1}(r)+c_{0,J-1}(r)\nonumber\\
&  +k^{2}\Big\{\frac{1}{2}r_{J-1}s_{0,J-1}(r)-\frac{1}{a_{J-1}}s_{2,J-1}%
(r)+c_{2,J-1}(r)\Big\}+\mathcal{O}(k^{4}),
\end{align}%
\begin{equation}
V_{\alpha}(r)=q_{0}c_{0,J+1}(r)+k^{2}\Big\{q_{1}c_{0,J+1}(r)+q_{0}%
c_{2,J+1}(r)\Big\}+\mathcal{O}(k^{4}),
\end{equation}%
\begin{align}
U_{\beta}(r)  &  =q_{0}\frac{1}{a_{J+1}}s_{0,J-1}(r)\nonumber\\
&  +k^{2}\Big\{q_{0}\frac{1}{a_{J+1}}s_{2,J-1}(r)-q_{0}\frac{r_{J+1}}%
{2}s_{0,J-1}(r)+q_{1}\frac{1}{a_{J+1}}s_{0,J-1}(r)\Big\}+\mathcal{O}(k^{4}),
\end{align}%
\begin{align}
V_{\beta}(r)  &  =\frac{-1}{a_{J+1}}s_{0,J+1}(r)+c_{0,J+1}(r)\nonumber\\
&  +k^{2}\Big\{\frac{1}{2}r_{J+1}s_{0,J+1}(r)-\frac{1}{a_{J+1}}s_{2,J+1}%
(r)+c_{2,J+1}(r)\Big\}+\mathcal{O}(k^{4}). \label{eqn:vbetwavefinal}%
\end{align}

As in the single channel case, the tool that we use to derive the causality
bound is the Wronskian identity. \ Through the derivation we assume that the potential is not singular at the origin, and that regular
solutions of the Schr\"{o}dinger equations $U(r)$ and $V(r)$ for two different values of momenta, $k_{a}$ and $k_{b}$, satisfy
\begin{equation}
\lim_{\rho\rightarrow0^{+}}U_{b}(\rho)U_{a}^{\prime}(\rho)=\lim_{\rho
\rightarrow0^{+}}U_{a}(\rho)U_{b}^{\prime}(\rho) = 0,
\end{equation}
\begin{equation}
\lim_{\rho\rightarrow0^{+}}V_{b}(\rho)V_{a}^{\prime}(\rho)=\lim_{\rho
\rightarrow0^{+}}V_{a}(\rho)V_{b}^{\prime}(\rho) = 0.
\end{equation}
For $\gamma=\alpha,\beta$ states we obtain
\begin{align}
(k_{a}^{2}-k_{b}^{2})  &  \int_{0}^{r}[U_{a\gamma}(r^{\prime})U_{b\gamma
}(r^{\prime})+V_{a\gamma}(r^{\prime})V_{b\gamma}(r^{\prime})]\,dr^{\prime
}\nonumber\label{eqn:sum2}\\
&  =W[U_{a\gamma}(r),U_{b\gamma}(r)]+W[V_{a\gamma}(r),V_{b\gamma}(r)],
\end{align}
and for the combination of $\alpha$ and $\beta$ states, we get
\begin{align}
(k_{a}^{2}-k_{b}^{2})  &  \int_{0}^{r}[U_{a\alpha}(r^{\prime})U_{b\beta
}(r^{\prime})+V_{a\alpha}(r^{\prime})V_{b\beta}(r^{\prime})+U_{b\alpha
}(r^{\prime})U_{a\beta}(r^{\prime})+V_{b\alpha}(r^{\prime})V_{a\beta
}(r^{\prime})]\,dr^{\prime}\nonumber\\
&  =W[U_{a\alpha}(r),U_{b\beta}(r)]+W[U_{a\beta}(r),U_{b\alpha}%
(r)]+W[V_{a\alpha}(r),V_{b\beta}(r)]+W[V_{a\beta}(r),V_{b\alpha}(r)].
\label{eqn:sum5}%
\end{align}
The Wronskian of the $\alpha$-state wave functions and the $\beta$-state wave
functions for the non-interacting region $r\geq R$ are given in Appendix
\ref{append:Wronskiansof functions}.

In Eq.~(\ref{eqn:sum2}), we set $k_{a}=0$ and take the limit $k=k_{b}%
\rightarrow0$. In the region $r\geq R$ we obtain the following relations for
the effective range parameters,
\begin{align}
&  r_{J-1}=b_{J-1}(r)+2q_{0}^{2}W[c_{2}(r),c_{0}(r)]_{J+1}-2\int_{0}%
^{r}\left(  \left[  U_{\alpha}^{(0)}(r^{\prime})\right]  ^{2}+\left[
V_{\alpha}^{(0)}(r^{\prime})\right]  ^{2}\right)  dr^{\prime}%
,\label{eqn:effectrangAlph2}\\
&  r_{J+1}=b_{J+1}(r)+2q_{0}^{2}\frac{1}{a_{J+1}^{2}}W[s_{2}(r),s_{0}%
(r)]_{J-1}-2\int_{0}^{r}\left(  \left[  U_{\beta}^{(0)}(r^{\prime}){}\right]
^{2}+\left[  V_{\beta}^{(0)}(r^{\prime})\right]  ^{2}\right)  dr^{\prime}.
\label{eqn:effectrangBet2}%
\end{align}
Here $b_{J\mp1}$ are
\begin{align}
b_{J\mp1}(r)  &  =\frac{2}{a_{J\mp1}^{2}}W[s_{2}(r),s_{0}(r)]_{J\mp1}+\frac
{2}{a_{J\mp1}}W[c_{0}(r),s_{2}(r)]_{J\mp1}\nonumber\\
&  +\frac{2}{a_{J\mp1}}W[s_{0}(r),c_{2}(r)]_{J\mp1}+2W[c_{2}(r),c_{0}%
(r)]_{J\mp1}, \label{eqn:bAlph1}%
\end{align}
which reduce to the form
\begin{align}
b_{J\mp1}(r)  &  =\frac{1}{a_{J\mp1}^{2}}\frac{2\pi}{\Gamma\left(  J\mp
1+\frac{3}{2}\right)  \Gamma\left(  J\mp1+\frac{5}{2}\right)  }\left(
\frac{r}{2}\right)  ^{2(J\mp1)+3}\nonumber\\
&  -\frac{1}{a_{J\mp1}}\frac{4}{J\mp1+\frac{1}{2}}\left(  \frac{r}{2}\right)
^{2}-\frac{2\Gamma\left(  J\mp1-\frac{1}{2}\right)  \Gamma\left(  J\mp
1+\frac{1}{2}\right)  }{\pi}\left(  \frac{r}{2}\right)  ^{-2(J\mp1)+1}.
\label{eqn:generalb}%
\end{align}

In Eq.~(\ref{eqn:sum5}), we set $k_{a}=0$ and take the same limit,
$k=k_{b}\rightarrow0$. In the region $r\geq R$ we obtain
\begin{equation}
q_{1}\frac{2}{a_{J+1}}=d_{J}(r)-2\int_{0}^{r}\left[  U_{\alpha}^{(0)}%
(r^{\prime})U_{\beta}^{(0)}(r^{\prime})+V_{\alpha}^{(0)}(r^{\prime})V_{\beta
}^{(0)}(r^{\prime})\right]  \,dr^{\prime}. \label{eqn:sum8}%
\end{equation}
Here $d_{J}(r)$ is
\begin{align}
d_{J}(r)=  &  -q_{0}\frac{2}{a_{J-1}a_{J+1}}W[s_{2}(r),s_{0}(r)]_{J-1}%
+2q_{0}W[c_{2}(r),c_{0}(r)]_{J+1}\nonumber\\
&  +q_{0}\frac{2}{a_{J+1}}\Big\{W[c_{2}(r),s_{0}(r)]_{J-1}-W[c_{2}%
(r),s_{0}(r)]_{J+1}\Big\}, \label{eqn:sum9}%
\end{align}
and this can be written as
\begin{align}
d_{J}(r)=  &  \frac{-q_{0}}{a_{J-1}a_{J+1}}\frac{2\pi}{\Gamma\left(  \frac
{1}{2}+J\right)  \Gamma\left(  \frac{3}{2}+J\right)  }\left(  \frac{r}%
{2}\right)  ^{2J+1}\nonumber\\
&  +\frac{q_{0}}{a_{J+1}}\frac{4}{(2J-1)(2J+3)}r^{2}-2q_{0}\frac{\Gamma\left(
J+\frac{1}{2}\right)  \Gamma\left(  J+\frac{3}{2}\right)  }{\pi}\left(
\frac{r}{2}\right)  ^{-2J-1}. \label{eqn:sum10}%
\end{align}

All of equations derived here have been numerically checked using a simple
potential model.\ The numerical calculations using
delta-function shell potentials with partial-wave mixing have been performed, and details are
given in Ref.~\cite{CausalityBounds2012V1}.

\section{Causality Bounds}

\label{sec:causalityBounds}

The terms in the integrals in Eq.~(\ref{eqn:effectrangAlph2}) and
Eq.~(\ref{eqn:effectrangBet2}) are positive semi-definite since the wave
functions are real. Therefore Eq.~(\ref{eqn:effectrangAlph2}) and
Eq.~(\ref{eqn:effectrangBet2}) place upper bounds for the effective range
$r_{J-1}$ and $r_{J+1}$ respectively. As noted in the introduction, these
upper bounds result from the causality and unitarity in the quantum scattering
problem. \ Our results are extensions of single-channel results in
Ref.~\cite{Phillips1997} for the $S$-wave in three dimensions and in
Ref.~\cite{Hammer-Lee9} for arbitrary angular momentum and arbitrary dimensions.

The causality bounds for the lower and higher partial-wave effective ranges
are
\begin{equation}
r_{J-1}\leq b_{J-1}(r)-2q_{0}^{2}\frac{\Gamma(J+\frac{1}{2})\Gamma(J+\frac
{3}{2})}{\pi}\left(  \frac{r}{2}\right)  ^{-2J-1}, \label{eqn:rJ-1bound}%
\end{equation}%
\begin{equation}
r_{J+1}\leq b_{J+1}(r)+\frac{2q_{0}^{2}}{a_{J+1}^{2}}\frac{\pi}{\Gamma
(J+\frac{1}{2})\Gamma(J+\frac{3}{2})}\left(  \frac{r}{2}\right)  ^{2J+1}.
\label{eqn:rJ+1bound}%
\end{equation}
We note that the effective range bounds are modified due to partial-wave
mixing. \ The causality upper bound for $r_{J-1}$ is lowered by the negative
term on the right hand side of Eq.~(\ref{eqn:rJ-1bound}), while the causality
upper bound for the higher partial-wave is increased by the term on the right
hand side of Eq.~(\ref{eqn:rJ+1bound}). \ When $q_{0}$ is nonzero and we take
the limit of zero range interactions, Eq.~(\ref{eqn:rJ-1bound}) tells us that
$r_{J-1}$ is driven to negative infinity for any $J$. \ We conclude that the
physics of partial-wave mixing requires a non-zero range for the interactions
in order to comply with the constraints of causality and unitarity. \ In Ref.
\cite{Hammer-Lee9} a similar negative divergence in the effective range
parameter was found for single-channel partial waves with $L>0$. \ What is
interesting here is that the negative divergence of the effective range occurs
already in the $^{3}S_{1}$ channel due to partial-wave mixing.

We note that the integral terms in Eq.~(\ref{eqn:effectrangAlph2}),
Eq.~(\ref{eqn:effectrangBet2}) and Eq.~(\ref{eqn:sum8}) are closely related.
Analysis of these equations using the Cauchy-Schwarz inequality provides
another useful relation for the coupled-channel wave functions. For real
functions $f_{1}(r)$, $f_{2}(r)$, $g_{1}(r)$ and $g_{2}(r)$, the
Cauchy-Schwarz inequality is%
\begin{align}
\Big(\int[f_{1}(r)\ f_{2}(r)]\left[
\begin{array}
[c]{c}%
f_{1}(r)\\
f_{2}(r)
\end{array}
\right]  \,dr\Big)\Big(\int[g_{1}(r)\ g_{2}(r)]  &  \left[
\begin{array}
[c]{c}%
g_{1}(r)\\
g_{2}(r)
\end{array}
\right]  \,dr\Big)\nonumber\\
\geq &  \Big|\int[f_{1}(r)g_{1}(r)+f_{2}(r)g_{2}(r)]\,dr\Big|^{2}.
\label{eqn:Cauch-Sch_Eq}%
\end{align}
When we apply the inequality to our coupled wave functions, we get
\begin{equation}
f_{J-1}(r)g_{J+1}(r)\geq\left[  h_{J}(r)\right]  ^{2},
\label{eqn:the Cauchy-Schwarz ineqality}%
\end{equation}
where
\begin{equation}
f_{J-1}(r)=b_{J-1}(r)-2q_{0}^{2}\frac{\Gamma(J+\frac{1}{2})\Gamma(J+\frac
{3}{2})}{\pi}\left(  \frac{r}{2}\right)  ^{-2J-1}-r_{J-1},
\label{eqn:f(r)CoupledChannel}%
\end{equation}%
\begin{equation}
g_{J+1}(r)=b_{J+1}(r)+\frac{2q_{0}^{2}}{a_{J+1}^{2}}\frac{\pi}{\Gamma
(J+\frac{1}{2})\Gamma(J+\frac{3}{2})}\left(  \frac{r}{2}\right)
^{2J+1}-r_{J+1}, \label{eqn:g(r)CoupledChannel}%
\end{equation}
and
\begin{equation}
h_{J}(r)=d_{J}(r)-q_{1}\frac{2}{a_{J+1}}. \label{eqn:h(r)CoupledChannel}%
\end{equation}
This inequality is used to define a Cauchy-Schwarz range, $R^{\text{C-S}}$, as
the minimum $r$ for each coupled channel where Eq.~(\ref{eqn:rJ-1bound}),
Eq.~(\ref{eqn:rJ+1bound}) and Eq.~(\ref{eqn:the Cauchy-Schwarz ineqality}) hold.

\section{Neutron-Proton Scattering}

\label{sec:Proton-Neutron-by-Nijmegen}

We now apply our causality bounds to physical neutron-proton data. \ In this
study, we use the low energy neutron-proton scattering data (0-350 MeV) from
the NN data base by the Nijmegen Group \cite{Nijmegen1993}. Table
\ref{table:tableforOnechannelWaves} - \ref{table:tableforquad} show the
low-energy threshold parameters in the eigenphase parameterization for the
NijmII and the Reid93 potentials. These parameters are calculated using the
results obtained in Ref. \cite{ValderramaArriola} for the low-energy threshold
parameters of the nuclear bar parameterization and relations between
eigenphase and nuclear bar parameterizations given in Appendix \ref{append:B}.
Using these numbers we analyze Eq.~(\ref{eqn:effectrangAlph2}),
Eq.~(\ref{eqn:effectrangBet2}) and Eq.~(\ref{eqn:sum8}), as well as causality
bounds for the uncoupled channels.

\begin{table}[tbh]
\caption{The eigenphase low energy parameters of uncoupled channels for
neutron-proton scattering by the NijmII and the Reid93 interaction
potentials.}%
\label{table:tableforOnechannelWaves}
\centering%
\begin{tabular}
[c]{||c|c|c||}\hline\hline
Channel & $a_{L}$ [$\text{fm}^{2L+1}$] & $r_{L}$ [$\text{fm}^{-2L+1}$]\\
& NijmII (Reid93) & NijmII (Reid93)\\\hline\hline
$^{1}S_{0}$ & -23.727 (-23.735) & 2.670 (2.753)\\\hline
$^{1}P_{1}$ & 2.797 (2.736) & -6.399 (-6.606)\\\hline
$^{3}P_{0}$ & -2.468 (-2.469) & 3.914 (3.870)\\\hline
$^{3}P_{1}$ & 1.529 (1.530) & -8.580 (-8.556)\\\hline
$^{1}D_{2}$ & -1.389 (-1.377) & 14.87 (15.04)\\\hline
$^{3}D_{2}$ & -7.405 (-7.411) & 2.858 (2.851)\\\hline
$^{1}F_{3}$ & 8.383 (8.365) & -3.924 (-3.936)\\\hline
$^{3}F_{3}$ & 2.703 (2.686) & -9.932 (-9.994)\\\hline\hline
\end{tabular}
\end{table}\begin{table}[tbh]
\caption{The eigenphase low energy parameters of coupled channels for
neutron-proton scattering by the NijmII and the Reid93 interaction
potentials.}%
\label{table:tableforTwochannelWaves}
\centering%
\begin{tabular}
[c]{||c|c|c||}\hline\hline
Channel & $a_{L}$ [$\text{fm}^{2L+1}$] & $r_{L}$ [$\text{fm}^{-2L+1}$]\\
& NijmII (Reid93) & NijmII (Reid93)\\\hline\hline
$^{3}S_{1}$ & 5.418 (5.422) & 1.7531 (1.7554)\\\hline
$^{3}D_{1}$ & 6.0043 (5.9539) & -3.523 (-3.566)\\\hline
$^{3}P_{2}$ & -0.2844 (-0.2892) & -11.1465 (-10.7127)\\\hline
$^{3}F_{2}$ & 8.126 (7.882) & -5.640 (-5.821)\\\hline
$^{3}D_{3}$ & -0.1449 (-0.177) & 288.428 (198.528)\\\hline
$^{3}G_{3}$ & 648.813 (534.594) & -0.03306 (-0.0529)\\\hline\hline
\end{tabular}
\end{table}\begin{table}[tbh]
\caption{The eigenphase low energy mixing parameters of coupled channels for
neutron-proton scattering by the NijmII and the Reid93 interaction
potentials.}%
\label{table:tableforquad}
\centering
\begin{tabular}
[c]{||c|c|c||}\hline\hline
Mixing angle & $q_{0}$ [$\text{fm}^{2}$] & $q_{1}$ [$\text{fm}^{4}$]\\
& NijmII (Reid93) & NijmII (Reid93)\\\hline\hline
$\varepsilon_{1}$ & 0.303987 (0.303394) & -2.00228 (-1.99129)\\\hline
$\varepsilon_{2}$ & -5.65752 (-5.5325) & 65.8602 (64.2979)\\\hline
$\varepsilon_{3}$ & 66.6632 (54.7062) & 340.988 (94.9015)\\\hline\hline
\end{tabular}
\end{table}

\subsection{\emph{Uncoupled}\textbf{\emph{\ Channels}}}

\label{sec:Cases without Mixing} We start with channels of a single uncoupled
partial wave. Since there is no mixing between different partial-waves, we
evaluate Eq.~(\ref{eqn:effectrangAlph2}) and Eq.~(\ref{eqn:effectrangBet2})
with zero mixing angle, and we obtain the following equation for the effective
range
\begin{equation}
r_{L}=b_{L}(r)-2\int_{0}^{r}\Big[U^{(0)}(r^{\prime})\Big]^{2}\,dr^{\prime},
\label{eqn:EffecRangNoMix}%
\end{equation}
where $b_{L}$ is given in Eq.~(\ref{eqn:uncoup:generalb}). These solutions
were derived by Hammer and Lee \cite{HammerDean27} for arbitrary dimension and
angular momentum.

\begin{figure}[ptb]
\begin{center}
\resizebox{100mm}{!}{\includegraphics{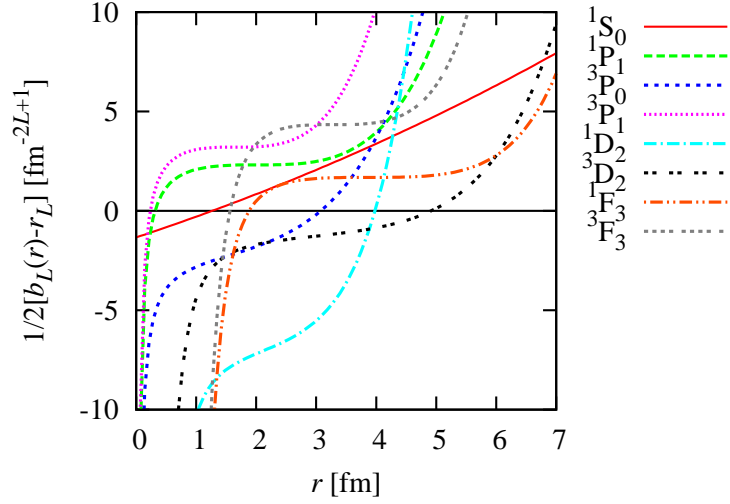}}
\end{center}
\caption{The plot of $[b_{L}(r)-r_{L}]/2$ as a function of $r$ for
neutron-proton scattering via the NijmII potential in the $^{2S+1}L_{J}$
channel.}%
\label{figure:SingletALL}%
\end{figure}

Here, we analyze the causality bound of the effective range for $L\leq3$ using
the scattering parameters in Table \ref{table:tableforOnechannelWaves}. In
Fig.~(\ref{figure:SingletALL}), we plot $\frac{1}{2}[b_{L}(r)-r_{L}]$ for all
of uncoupled channels with $L\leq3$. \ The physical region corresponds with
$\frac{1}{2}[b_{L}(r)-r_{L}]\geq0$.

For \emph{S}-wave scattering
\begin{equation}
b_{0}(r)=\frac{2}{3a_{0}^{2}}r^{3}-\frac{2}{a_{0}}r^{2}+2r,
\label{eqn:S_wave_b}%
\end{equation}
for \emph{P}-wave,
\begin{equation}
b_{1}(r)=\frac{2r^{5}}{45a_{1}^{2}}-\frac{2r^{2}}{3a_{1}}-\frac{2}{r},
\label{eqn:P_wave_b}%
\end{equation}
for \emph{D}-wave,
\begin{equation}
b_{2}(r)=\frac{2}{1575a_{2}^{2}}r^{7}-\frac{2}{5a_{2}}r^{2}-\frac{6}{r^{3}},
\label{eqn:D_wave_b}%
\end{equation}
for \emph{F}-wave,
\begin{equation}
b_{3}(r)=\frac{2r^{9}}{99225a_{3}^{2}}-\frac{2r^{2}}{7a_{3}}-\frac{90}{r^{5}},
\label{eqn:F_wave_b}%
\end{equation}
and for \emph{G}-wave,
\begin{equation}
b_{4}(r)=\frac{2r^{11}}{9823275a_{4}^{2}}-\frac{2r^{2}}{9a_{4}}-\frac
{3150}{r^{7}}. \label{eqn:G_wave_b}%
\end{equation}

\subsection{\textbf{\emph{Coupled Channels}}}

\label{sec:triplet Channel} We now analyze channels with coupled partial
waves. \ We plot Eq.~(\ref{eqn:f(r)CoupledChannel}) and
Eq.~(\ref{eqn:g(r)CoupledChannel}) for all coupled channels with $J\leq3$.
\ The physical region correspond both $f_{J-1}(r)\geq0$ and $g_{J+1}(r)\geq0$.
\begin{figure}[ptb]
\begin{center}
\resizebox{100mm}{!}{\includegraphics{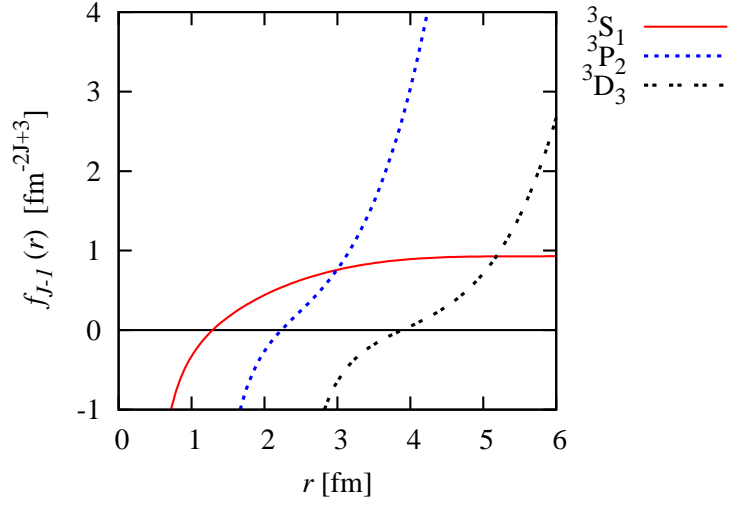}}
\end{center}
\caption{The plot of $f_{J-1}(r)$ as a function of $r$ for neutron-proton
scattering via the NijmII potential for $J\leq3$. Here $f_{1}(r)$ is rescaled
by a factor of $0.01$ and $f_{2}(r)$ is rescaled by a factor of $10^{-4}$.}%
\label{figure:AllTripletLOWER}%
\end{figure}\begin{figure}[ptb]
\begin{center}
\resizebox{100mm}{!}{\includegraphics{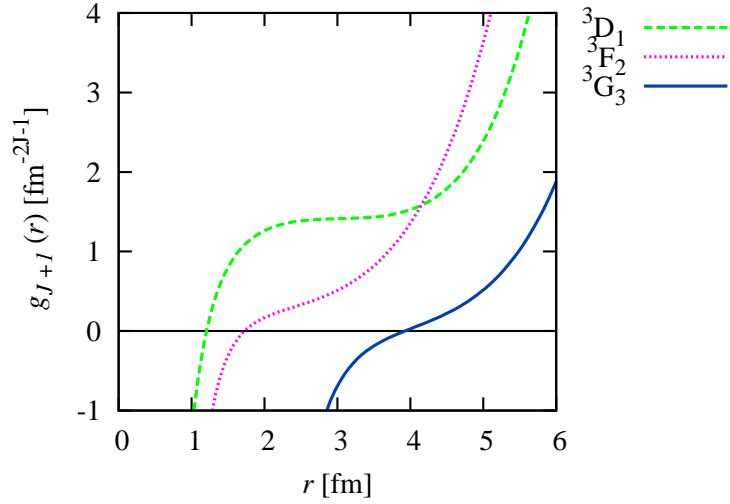}}
\end{center}
\caption{The plot of $g_{J+1}(r)$ as a function of $r$ for neutron-proton
scattering via the NijmII potential for $J\leq3$. Here $g_{3}(r)$ is rescaled
by a factor of $0.1$.}%
\label{figure:AllTripletHIGHER}%
\end{figure}

\subsubsection{\textbf{$\, ^{3}$S$_{1}$-$\, ^{3}$D$_{1}$ Coupling.}}

\label{sec:S-D Proton-Neutron}

We now consider Eq.~(\ref{eqn:effectrangAlph2}) - (\ref{eqn:sum9}) for the
$^{3}S_{1}$-$^{3}D_{1}$ coupled channel. We evaluate the Wronskians for $J=1$
and get
\begin{equation}
b_{0}(r)-q_{0}^{2}\frac{6}{r^{3}}-r_{0}=2\int_{0}^{r}\left(  \left[
U_{\alpha}^{(0)}(r^{\prime})\right]  ^{2}+ \left[  V_{\alpha}^{(0)}(r^{\prime
})\right]  ^{2} \right)  dr^{\prime}, \label{eqn:r3S1}%
\end{equation}%
\begin{equation}
b_{2}(r)+q_{0}^{2}\frac{2r^{3}}{3a_{2}^{2}}-r_{2}=2\int_{0}^{r}\left(  \left[
U_{\beta}^{(0)}(r^{\prime})\right]  ^{2} + \left[  V_{\beta}^{(0)}(r^{\prime
})\right]  ^{2} \right)  dr^{\prime}, \label{eqn:r3D1}%
\end{equation}%
\begin{equation}
d_{1}(r)-q_{1}\frac{2}{a_{2}}=2\int_{0}^{r}[U_{0\alpha}(r^{\prime})U_{0\beta
}(r^{\prime})+V_{0\alpha}(r^{\prime})V_{0\beta}(r^{\prime})]\,dr^{\prime}.
\label{eqn:q1 boundE1}%
\end{equation}
$b_{0}(r)$ and $b_{2}(r)$ are given in Eq.~(\ref{eqn:S_wave_b}) and in
Eq.~(\ref{eqn:D_wave_b}), respectively, and $d_{1}(r)$ is
\begin{equation}
d_{1}(r)=-q_{0}\frac{1}{a_{0}a_{2}}\frac{2r^{3}}{3}+q_{0}\frac{1}{a_{2}}%
\frac{4r^{2}}{5}-q_{0}\frac{6}{r^{3}}. \label{eqn:d1 bound}%
\end{equation}

\begin{figure}[ptb]
\begin{center}
\resizebox{80mm}{!}{\includegraphics{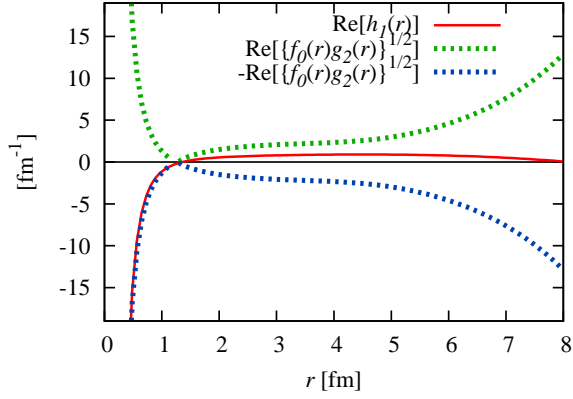}}
\end{center}
\caption{We plot Re$\left[  \sqrt{f_{0}(r)g_{2}(r)}\right]  $, $-$Re$\left[
\sqrt{f_{0}(r)g_{2}(r)}\right]  $, and Re$\left[  h_{1}(r)\right]  $ as
functions of $r$ for neutron-proton scattering in $^{3}S_{1}$-$^{3}D_{1}$
coupled channel.}%
\label{fig:figureS_D_MixingNijmII}%
\end{figure}

Using the scattering parameters in Table \ref{table:tableforTwochannelWaves} -
\ref{table:tableforquad}, we plot Eq.~(\ref{eqn:r3S1}), Eq.~(\ref{eqn:r3D1})
and Eq.~(\ref{eqn:q1 boundE1}) as functions of $r$. In
Fig.~(\ref{fig:figureS_D_MixingNijmII}) we show the physical region where the
causality bounds $f_{0}(r)\geq0$, $g_{2}(r)\geq0$, and $f_{0}(r)g_{2}(r)\geq
h_{1}^{2}(r)$, are satisfied. \ Here we have%
\begin{equation}
f_{0}(r)=\frac{2}{3a_{0}^{2}}r^{3}-\frac{2}{a_{0}}r^{2}+2r-q_{0}^{2}\frac
{6}{r^{3}}-r_{0}, \label{eqn:f_0(r)}%
\end{equation}%
\begin{equation}
g_{2}(r)=\frac{2}{1575a_{2}^{2}}r^{7}-\frac{2}{5a_{2}}r^{2}-\frac{6}{r^{3}%
}+q_{0}^{2}\frac{2r^{3}}{3a_{2}^{2}}-r_{2}, \label{eqn:g_2(r)}%
\end{equation}%
\begin{equation}
h_{1}(r)=-q_{0}\frac{1}{a_{0}a_{2}}\frac{2r^{3}}{3}+q_{0}\frac{1}{a_{2}}%
\frac{4r^{2}}{5}-q_{0}\frac{6}{r^{3}}-q_{1}\frac{2}{a_{2}}. \label{eqn:h_1(r)}%
\end{equation}

\subsubsection{\textbf{$\, ^{3}$P$_{2} -\, ^{3}$F$_{2}$ Coupling.}}

\label{sec:P-F Proton-Neutron}

\begin{figure}[ptb]
\begin{center}
\resizebox{80mm}{!}{\includegraphics{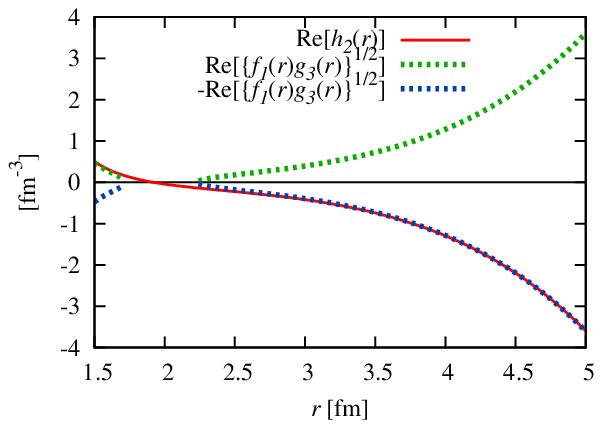}}
\end{center}
\caption{We plot Re$\left[  \sqrt{f_{1}(r)g_{3}(r)}\right]  $, $-$Re$\left[
\sqrt{f_{1}(r)g_{3}(r)}\right]  $, and Re$\left[  h_{2}(r)\right]  $ as
functions of $r$ for neutron-proton scattering in $^{3}P_{2}$-$^{3}F_{2}$
coupled channel. The functions are rescaled by a factor of $0.01$. }%
\label{fig:figureP_F_MixingNijmII}%
\end{figure}

In the $^{3}P_{2}$-$^{3}F_{2}$ coupled channel Eq.~(\ref{eqn:effectrangAlph2})
- (\ref{eqn:sum9}) take the following forms,
\begin{equation}
b_{1}(r)-q_{0}^{2}\frac{90}{r^{5}}-r_{1}=2\int_{0}^{r}\left(  \left[
U_{\alpha}^{(0)}(r^{\prime})\right]  ^{2}+ \left[  V_{\alpha}^{(0)}(r^{\prime
})\right]  ^{2} \right)  dr^{\prime}, \label{eqn:r3P2}%
\end{equation}%
\begin{equation}
b_{3}(r)+q_{0}^{2}\frac{1}{a_{3}^{2}}\frac{2r^{5}}{45}-r_{3}=2\int_{0}%
^{r}\left(  \left[  U_{\beta}^{(0)}(r^{\prime})\right]  ^{2} + \left[
V_{\beta}^{(0)}(r^{\prime})\right]  ^{2} \right)  dr^{\prime},
\label{eqn:r3F2}%
\end{equation}%
\begin{equation}
d_{2}(r)-q_{1}\frac{2}{a_{3}}=2\int_{0}^{r}[U_{0\alpha}(r^{\prime})U_{0\beta
}(r^{\prime})+V_{0\alpha}(r^{\prime})V_{0\beta}(r^{\prime})]\,dr^{\prime}.
\label{eqn:q1 boundE2}%
\end{equation}
$b_{1}(r)$ and $b_{3}(r)$ are defined in Eq.~(\ref{eqn:P_wave_b}) and
Eq.~(\ref{eqn:F_wave_b}), respectively, and $d_{2}(r)$ is
\begin{equation}
d_{2}(r)=-q_{0}\frac{1}{a_{1}a_{3}}\frac{2r^{5}}{45}+q_{0}\frac{1}{a_{3}}%
\frac{4r^{2}}{21}-q_{0}\frac{90}{r^{5}}. \label{eqn:eqn d2}%
\end{equation}
The causality bounds are $f_{1}(r)\geq0$, $g_{3}(r)\geq0$, and $f_{1}%
(r)g_{3}(r)\geq h_{2}^{2}(r)$, where
\begin{equation}
f_{1}(r)=\frac{2r^{5}}{45a_{1}^{2}}-\frac{2r^{2}}{3a_{1}}-\frac{2}{r}%
-q_{0}^{2}\frac{90}{r^{5}}-r_{1}, \label{eqn:f_1(r)}%
\end{equation}%
\begin{equation}
g_{3}(r)=\frac{2r^{9}}{99225a_{3}^{2}}-\frac{2r^{2}}{7a_{3}}-\frac{90}{r^{5}%
}+q_{0}^{2}\frac{1}{a_{3}^{2}}\frac{2r^{5}}{45}-r_{3}, \label{eqn:g_3(r)}%
\end{equation}%
\begin{equation}
h_{2}(r)=-q_{0}\frac{1}{a_{1}a_{3}}\frac{2r^{5}}{45}+q_{0}\frac{1}{a_{3}}%
\frac{4r^{2}}{21}-q_{0}\frac{90}{r^{5}}-q_{1}\frac{2}{a_{3}}.
\label{eqn:h_2(r)}%
\end{equation}
In Fig.~(\ref{fig:figureP_F_MixingNijmII}) we show the physical region for the
$^{3}P_{2}$-$^{3}F_{2}$ coupled channel wave functions. 

\subsubsection{\textbf{$\, ^{3}$D$_{3} -\, ^{3}$G$_{3}$ Coupling.}}

\label{sec:D-G Proton-Neutron}

\begin{figure}[ptb]
\begin{center}
\resizebox{80mm}{!}{\includegraphics{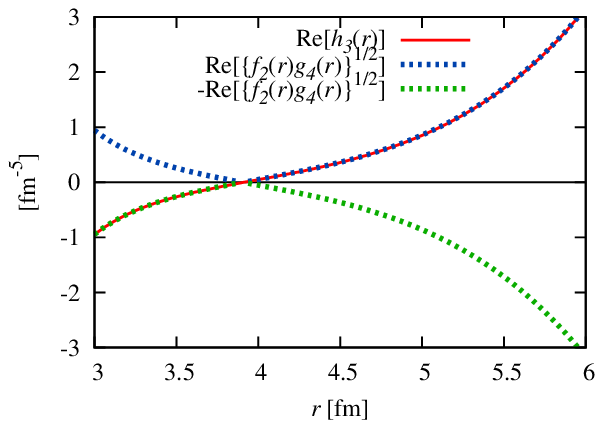}}
\end{center}
\caption{We plot Re$\left[  \sqrt{f_{2}(r)g_{4}(r)}\right]  $, $-$Re$\left[
\sqrt{f_{2}(r)g_{4}(r)}\right]  $, and Re$\left[  h_{3}(r)\right]  $ as
functions of $r$ for neutron-proton scattering in $^{3}D_{3}$-$^{3}G_{3}$
coupled channel. The functions are rescaled by a factor of $0.01$. }%
\label{fig:figureD_G_MixingNijmII}%
\end{figure}

For $J=3$, the $^{3}D_{3}$ and $^{3}G_{3}$ channels are coupled. In this case
Eq.~(\ref{eqn:effectrangAlph2})-(\ref{eqn:sum9}) read
\begin{equation}
b_{2}(r)-q_{0}^{2}\frac{3150}{r^{7}}-r_{2}=2\int_{0}^{r}\left(  \left[
U_{\alpha}^{(0)}(r^{\prime})\right]  ^{2}+ \left[  V_{\alpha}^{(0)}(r^{\prime
})\right]  ^{2} \right)  dr^{\prime}, \label{eqn:r3D3}%
\end{equation}%
\begin{equation}
b_{4}(r)+\frac{q_{0}^{2}}{a_{4}^{2}}\frac{2r^{7}}{1575}-r_{4}=2\int_{0}%
^{r}\left(  \left[  U_{\beta}^{(0)}(r^{\prime})\right]  ^{2} + \left[
V_{\beta}^{(0)}(r^{\prime})\right]  ^{2} \right)  dr^{\prime},
\label{eqn:r3G3}%
\end{equation}%
\begin{equation}
d_{3}(r)-q_{1}\frac{2}{a_{4}}=2\int_{0}^{r}[U_{0\alpha}(r^{\prime})U_{0\beta
}(r^{\prime})+V_{0\alpha}(r^{\prime})V_{0\beta}(r^{\prime})]\,dr^{\prime}.
\label{eqn:q1 boundE3}%
\end{equation}
Here $d_{3}(r)$ is
\begin{equation}
d_{3}(r)=-\frac{q_{0}}{a_{2}a_{4}}\frac{2r^{7}}{1575}+q_{0}\frac{1}{a_{4}%
}\frac{4r^{2}}{45}-q_{0}\frac{3150}{r^{7}}. \label{eqn:eqn d3}%
\end{equation}
The causality bounds are again $f_{2}(r)\geq0$, $g_{4}(r)\geq0$, and
$f_{2}(r)g_{4}(r)\geq h_{3}^{2}(r)$, where
\begin{equation}
f_{2}(r)=\frac{2r^{7}}{1575a_{2}^{2}}-\frac{2r^{2}}{5a_{2}}-\frac{6}{r^{3}%
}-q_{0}^{2}\frac{3150}{r^{7}}-r_{2}, \label{eqn:f_2(r)}%
\end{equation}%
\begin{equation}
g_{4}(r)=\frac{2r^{11}}{9823275a_{4}^{2}}-\frac{2r^{2}}{9a_{4}}-\frac
{3150}{r^{7}}+\frac{q_{0}^{2}}{a_{4}^{2}}\frac{2r^{7}}{1575}-r_{4},
\label{eqn:g_4(r)}%
\end{equation}%
\begin{equation}
h_{3}(r)=-\frac{q_{0}}{a_{2}a_{4}}\frac{2r^{7}}{1575}+q_{0}\frac{1}{a_{4}%
}\frac{4r^{2}}{45}-q_{0}\frac{3150}{r^{7}}-q_{1}\frac{2}{a_{4}}.
\label{eqn:h_3(r)}%
\end{equation}
We show plots for the $^{3}D_{3}$-$^{3}G_{3}$ channel in
Fig.~(\ref{fig:figureD_G_MixingNijmII}). 

\section{Results and Discussion}

In this section we present the results for the causal and Cauchy-Schwarz
ranges, $R^{b}$, and $R^{\text{C-S}}$. \ We use the NijmII\ scattering data
for neutron-proton scattering presented above. \ In Table
\ref{table:RcausalSingleChannel} we show results for the causal range for all
uncoupled channels by setting
\begin{equation}
r_{L}=b_{L}(r).
\end{equation}
In Table \ref{table:RcausalCoupledChannel} we determine the causal range for
all coupled channels using Eq.~(\ref{eqn:f(r)CoupledChannel}%
)-(\ref{eqn:g(r)CoupledChannel}). \ Also, we find the Cauchy-Schwarz ranges
shown in Table \ref{table:R_cauchy_schwarz} using
Eq.~(\ref{eqn:the Cauchy-Schwarz ineqality}). \begin{table}[tbh]
\caption{The causal ranges for uncoupled channels.}%
\label{table:RcausalSingleChannel}
\centering%
\begin{tabular}
[c]{||c|c|c|c|c|c|c|c|c||}\hline\hline
Channels & $^{1}S_{0}$ & $^{1}P_{1}$ & $^{3}P_{0}$ & $^{3}P_{1}$ & $^{1}D_{2}
$ & $^{3}D_{2}$ & $^{1}F_{3}$ & $^{3}F_{3}$\\\hline\hline
$R^{b}$ [fm] & $\ 1.27\ $ & $\ 0.31\ $ & $\ 3.07\ $ & $\ 0.23\ $ & $\ 3.98\ $
& $\ 4.91\ $ & $\ 1.88\ $ & $\ 1.56$\\\hline\hline
\end{tabular}
\end{table}\begin{table}[tbh]
\caption{The causal ranges for coupled channels.}%
\label{table:RcausalCoupledChannel}
\centering%
\begin{tabular}
[c]{||c|c|c|c|c|c|c||}\hline\hline
Channels & $^{3}S_{1}$ & $^{3}D_{1}$ & $^{3}P_{2}$ & $^{3}F_{2}
$ & $^{3}D_{3}$ & $^{3}G_{3}$\\\hline\hline
$R^{b}$ [fm] & $\ 1.29\ $ & $\ 1.20\ $ & $\ 2.23\ $& $\ 1.73\ $ & $\ 4.03\ $
& $\ 3.92$\\\hline\hline
\end{tabular}
\end{table}\begin{table}[tbh]
\caption{The Cauchy-Schwarz ranges for coupled channels.}%
\label{table:R_cauchy_schwarz}
\centering%
\begin{tabular}
[c]{||c|c|c|c||}\hline\hline
Channels & \ $^{3}S_{1}$-$^{3}D_{1}$ \  & \ $^{3}P_{2}$-$^{3}F_{2}$ \  &
\ $^{3}D_{3}$-$^{3}G_{3}$ \ \\\hline\hline
$R^{\text{C-S}}$ [fm] & $\ 1.29\ $ & $\ 4.65\ $ & $\ 5.68\ $\\\hline\hline
\end{tabular}
\end{table}\ We find that in some channels the causal and Cauchy-Schwarz
ranges are surprisingly large, and it is worthwhile to probe the origin of
these large ranges.

It is convenient to collect together some of the key formulas derived above.
\ The Cauchy-Schwarz inequality has the form
\begin{equation}
f_{J-1}(r)g_{J+1}(r)\geq\left[  h_{J}(r)\right]  ^{2}, \label{CauchySchwarz2}%
\end{equation}
where%
\begin{equation}
f_{J-1}(r)=b_{J-1}(r)-2q_{0}^{2}\frac{\Gamma(J+\frac{1}{2})\Gamma(J+\frac
{3}{2})}{\pi}\left(  \frac{r}{2}\right)  ^{-2J-1}-r_{J-1},
\end{equation}%
\begin{equation}
g_{J+1}(r)=b_{J+1}(r)+\frac{2q_{0}^{2}}{a_{J+1}^{2}}\frac{\pi}{\Gamma
(J+\frac{1}{2})\Gamma(J+\frac{3}{2})}\left(  \frac{r}{2}\right)
^{2J+1}-r_{J+1},
\end{equation}%
\begin{equation}
h_{J}(r)=d_{J}(r)-q_{1}\frac{2}{a_{J+1}},
\end{equation}%
\begin{align}
b_{J\mp1}(r)  &  =\frac{1}{a_{J\mp1}^{2}}\frac{2\pi}{\Gamma\left(  J\mp
1+\frac{3}{2}\right)  \Gamma\left(  J\mp1+\frac{5}{2}\right)  }\left(
\frac{r}{2}\right)  ^{2(J\mp1)+3}\nonumber\\
&  -\frac{1}{a_{J\mp1}}\frac{4}{J\mp1+\frac{1}{2}}\left(  \frac{r}{2}\right)
^{2}-\frac{2\Gamma\left(  J\mp1-\frac{1}{2}\right)  \Gamma\left(  J\mp
1+\frac{1}{2}\right)  }{\pi}\left(  \frac{r}{2}\right)  ^{-2(J\mp1)+1},
\end{align}%
\begin{align}
d_{J}(r)=  &  \frac{-q_{0}}{a_{J-1}a_{J+1}}\frac{2\pi}{\Gamma\left(  \frac
{1}{2}+J\right)  \Gamma\left(  \frac{3}{2}+J\right)  }\left(  \frac{r}%
{2}\right)  ^{2J+1}\nonumber\\
&  +\frac{q_{0}}{a_{J+1}}\frac{4}{(2J-1)(2J+3)}r^{2}-2q_{0}\frac{\Gamma\left(
J+\frac{1}{2}\right)  \Gamma\left(  J+\frac{3}{2}\right)  }{\pi}\left(
\frac{r}{2}\right)  ^{-2J-1}.
\end{align}
\bigskip

We note that the leading power of $r$ in $g_{J+1}(r)$ is%
\begin{equation}
\frac{1}{a_{J+1}^{2}}\frac{2\pi}{\Gamma\left(  J+1+\frac{3}{2}\right)
\Gamma\left(  J+1+\frac{5}{2}\right)  }\left(  \frac{r}{2}\right)  ^{2J+5}.
\end{equation}
This has a very small numerical prefactor multiplying $a_{J+1}^{-2}r^{2J+5}$.
\ For $J=1$ the factor is $2/1575$, for $J=2$ it is $2/99225$, and for $J=3 $
it is $2/9823275$. \ Therefore the term is negligible unless $r$ is large
compared with $(a_{J+1})^{1/(2J+3)}$. \ If we neglect this term, then the term
with the leading power of $r$ on the left hand side of
Eq.~(\ref{CauchySchwarz2}) is the same as that on the right hand side,%
\begin{align}
&  \frac{1}{a_{J-1}^{2}}\frac{2\pi}{\Gamma\left(  J-1+\frac{3}{2}\right)
\Gamma\left(  J-1+\frac{5}{2}\right)  }\left(  \frac{r}{2}\right)
^{2J+1}\cdot\frac{2q_{0}^{2}}{a_{J+1}^{2}}\frac{\pi}{\Gamma(J+\frac{1}%
{2})\Gamma(J+\frac{3}{2})}\left(  \frac{r}{2}\right)  ^{2J+1}\nonumber\\
&  =\left[  \frac{-q_{0}}{a_{J-1}a_{J+1}}\frac{2\pi}{\Gamma\left(  \frac{1}%
{2}+J\right)  \Gamma\left(  \frac{3}{2}+J\right)  }\left(  \frac{r}{2}\right)
^{2J+1}\right]  ^{2}.
\end{align}
As a result the curves for $f_{J-1}(r)g_{J+1}(r)$ and $\left[  h_{J}%
(r)\right]  ^{2}$ are approximately parallel for large $r$ until the term that
we have neglected becomes significant. \ These nearly parallel trajectories
inflate the value of the Cauchy-Schwarz range $r=R^{\text{C-S}}$ where the two
curves cross.

For the $^{3}S_{1}$-$^{3}D_{1}$ coupled channel we find $R^{\text{C-S}}$ is
about the same size as the Compton wavelength of the pion, $m_{\pi}^{-1}=1.5 $
fm. \ This is also comparable to what one expects for the range of the
nucleon-nucleon interaction. \ However the results are more interesting for
$J>1$. \ In the $^{3}P_{2}$-$^{3}F_{2}$ channel we have $R^{\text{C-S}}=4.65$
fm. \ And for the $^{3}D_{3}$-$^{3}G_{3}$ coupled channel we find
$R^{\text{C-S}}$ $=$ $5.68$ fm. \ These values are surprisingly large in
comparison with $m_{\pi}^{-1}$.

\section{One-Pion Exchange Potential}

We note that there are some channels where the causal range $R^{b}$ is also
quite large. \ By definition $R^{\text{C-S}}\geq R^{b}$ and so the
Cauchy-Schwarz range will then also be large. \ The causal range is the
minimum value for $r$ such that%
\begin{equation}
f_{J-1}(r)=b_{J-1}(r)-2q_{0}^{2}\frac{\Gamma(J+\frac{1}{2})\Gamma(J+\frac
{3}{2})}{\pi}\left(  \frac{r}{2}\right)  ^{-2J-1}-r_{J-1}\geq0
\end{equation}
for the lower partial wave, or%
\begin{equation}
g_{J+1}(r)=b_{J+1}(r)+\frac{2q_{0}^{2}}{a_{J+1}^{2}}\frac{\pi}{\Gamma
(J+\frac{1}{2})\Gamma(J+\frac{3}{2})}\left(  \frac{r}{2}\right)
^{2J+1}-r_{J+1}\geq0
\end{equation}
for the higher partial wave. \ For uncoupled channels we take $q_{0}=0$.

The largest values for $R^{b}$ occur when the effective range parameter is
positive or near zero. \ See for example the causal ranges for the $^{1}D_{2}
$, $^{3}D_{2}$, $^{3}D_{3}$, and $^{3}G_{3}$ channels. What happens is that
the function $f_{J-1}(r)$ or $g_{J+1}(r)$ remains negative with a rather small
slope until $r$ becomes quite large. \ The small slope is again associated
with the fact that the term with the highest power of $r$ has a small
numerical prefactor.

The range of the interaction plays the dominant role in setting the causal range. \ In the language of local potentials, this is the radius at which the magnitude of the potential is numerically very small. \ However there is also some influence of the exponential tail of the potential upon the causal range.

In all channels where the causal range is unusually large, $^{1}D_{2}$, $^{3}D_{2}$, $^{3}D_{3}$, and $^{3}G_{3}$, we find that the tail of the one-pion exchange potential is attractive. \ At smaller radii, the potential crosses over at some classical turning point to become
repulsive. \ See for example Fig. 2-4 in Ref. \cite{Stoks1993NijII}.

The detailed mechanism requires further study, but it appears that this geometry can cause a near-threshold wavepacket to reflect before reaching the classical turning point, thus mimicking a longer range potential. \ However some fine tuning is needed to produce a large causal range, as there is no enhancement in the $^{1}S_{0}$ and $^{3}S_{1}$ channels and a smaller amount of enhancement in the $^{3}P_{0}$ channel.

There seems to be no such enhancement of the causal range
in the $^{1}P_{1}$, $^{3}P_{1}$, and $^{3}D_{1}$ channels where the tail of the potential is repulsive. \ In fact, the causal range for the $^{1}P_{1}$ and $^{3}P_{1}$ channels are unusually small. \ This appears be related to
quantum tunneling into the inner region where the potential is attractive.

In the following analysis we will investigate the
importance of the tail of the one-pion exchange potential plays in setting the
causal range, $R^{b}$. \ We show that even though the one-pion exchange
potential is numerically small at distances larger than $5\,\text{fm}$,
chopping off the one-pion exchange tail at such distances produces a
non-negligible effect. \ The one-pion exchange potential tail appears to be
the source of the large values for $R^{b}$ in higher partial waves where the
central one-pion exchange tail is attractive.

\begin{figure}[ptb]
\begin{center}
\resizebox{100mm}{!}{\includegraphics{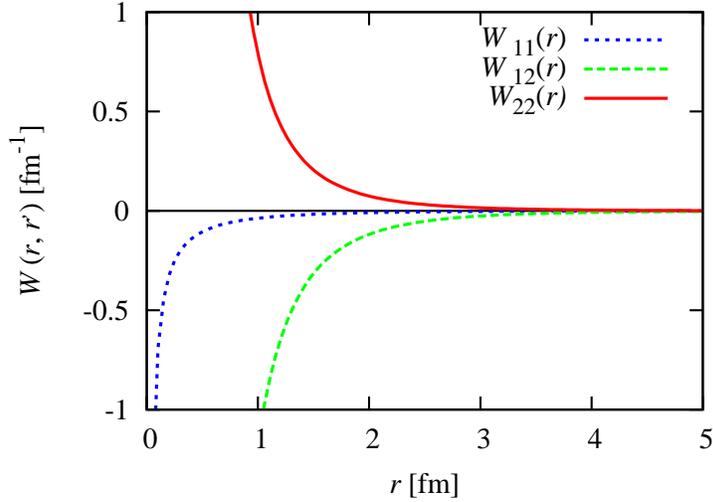}}
\end{center}
\caption{Plot of the potential matrix elements $W_{11}(r)=V_{C}(r)$,
$W_{12}(r)=W_{21}(r)=\sqrt{8}V_{T}(r)$ and $W_{22}(r)=V_{C}(r)-2V_{T}(r)$ as a
function of $r$ in the $^{3}S_{1}$-$^{3}D_{1}$ coupled channel.}%
\label{fig:WrR}%
\end{figure}

If we neglect electromagnetic effects, then the neutron-proton interaction
potential at long distances is governed by the one-pion exchange (OPE)
potential, which in configuration space is
\begin{equation}
V_{OPE}(r)=V_{C}(r)+S_{12}V_{T}(r).
\end{equation}
Here $V_{C}(r)$ is the central potential,
\begin{equation}
V_{C}(r)=\frac{g_{\pi N}^{2}}{12\pi}\left(  \frac{m_{\pi}}{2M_{N}}\right)
^{2}(\vec{\tau}_{1}\cdot\vec{\tau}_{2})(\vec{\sigma}_{1}\cdot\vec{\sigma}%
_{2})\frac{e^{-m_{\pi}r}}{r},\label{central potential}%
\end{equation}
$V_{T}(r)$ is the tensor potential,
\begin{equation}
V_{T}(r)=\frac{g_{\pi N}^{2}}{12\pi}\left(  \frac{m_{\pi}}{2M_{N}}\right)
^{2}\ (\vec{\tau}_{1}\cdot\vec{\tau}_{2})\ \Big(1+\frac{3}{m_{\pi}r}+\frac
{3}{(m_{\pi}r)^{2}}\Big)\frac{e^{-m_{\pi}r}}{r},
\end{equation}
and $S_{12}$ is the tensor operator,
\begin{equation}
S_{12}=3(\vec{\sigma_{1}}\cdot\hat{r})(\vec{\sigma_{2}}\cdot\hat{r}%
)-\vec{\sigma_{1}}\cdot\vec{\sigma_{2}}.\label{eqn:S12}%
\end{equation}
Here $m_{\pi}$ is the pion mass, $M_{N}$ is the nucleon mass, and $g_{\pi
N}=13.0$ is the pion-nucleon coupling constant. \ The one-pion exchange
potential is local in space and the interaction matrix in
Eq.~(\ref{eqn:PotentialMatrix}) takes the following form for $J=1$,%
\begin{equation}
W(r,r^{\prime})=\left(
\begin{array}
[c]{cc}%
V_{C}(r) & \sqrt{8}V_{T}(r)\\
\sqrt{8}V_{T}(r) & V_{C}(r)-2V_{T}(r)
\end{array}
\right)  \delta(r-r^{\prime}).\label{eqn:RealPotential}%
\end{equation}
In Fig.~(\ref{fig:WrR}) we plot $W_{11}(r)=V_{C}(r)$, $W_{12}(r)=W_{21}%
(r)=\sqrt{8}V_{T}(r),$ and $W_{22}(r)=V_{C}(r)-2V_{T}(r)$ in the $^{3}S_{1}%
$-$^{3}D_{1}$ coupled channel. 

To demonstrate the origin of large causal ranges found in Table
\ref{table:RcausalSingleChannel} and Table \ref{table:RcausalCoupledChannel},
we will present some simple but illustrative numerical examples. \ For each
channel we add a short range potential to the one-pion exchange potential in
order to reproduce the physical low-energy scattering parameters. The specific
model we use for the short range potential is not important to our general
analysis nor is it the most economical. \ We choose a simple scheme which
consists of three well-defined functions in three different regions and which
is continuously differentiable everywhere. \ The potential has the form%
\begin{align}
V(r)  &  =V_{\text{Gauss}}(r)\theta(R_{\text{Gauss}}-r)+V_{\text{Spline}%
}(r)\theta(r-R_{\text{Gauss}})\theta(R_{\text{Exch.}}-r)\nonumber\\
&  +V_{\text{Exch.}}(r)\theta(r-R_{\text{Exch.}}),
\end{align}
where $\theta$ is a unit step function.

\begin{figure}[ptb]
\begin{center}
\resizebox{100mm}{!}{\includegraphics{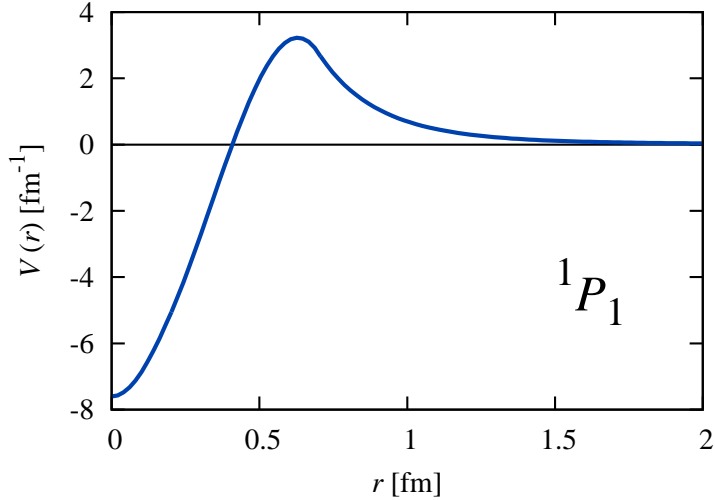}}
\end{center}
\caption{Plot of the model potential in the $^{1}P_{1}$ channel. In this
channel $S=0$, $T=0$, $C_{\pi}=9$, $C_{A}=405.9$, $C_{B}=769.5$, $C_{\Pi}=0$,
$C_{D}=0$, $C_{F}=0$, $C_{G}=-7.6$, $m_{A}=10.0m_{\pi}$, $m_{B}=5.45m_{\pi}$,
$m_{G}=4.46 m_{\pi}$, $C_{1}=-7.464$, $C_{2}=0.179$, $C_{3}=73.933$ and
$C_{4}=-78.011$. We use $R_{\text{Gauss}}=0.2$ fm and $R_{\text{Exch.}}=0.7$
fm.}%
\label{fig:OPE1P1}%
\end{figure}

\begin{figure}[ptb]
\begin{center}
\resizebox{100mm}{!}{\includegraphics{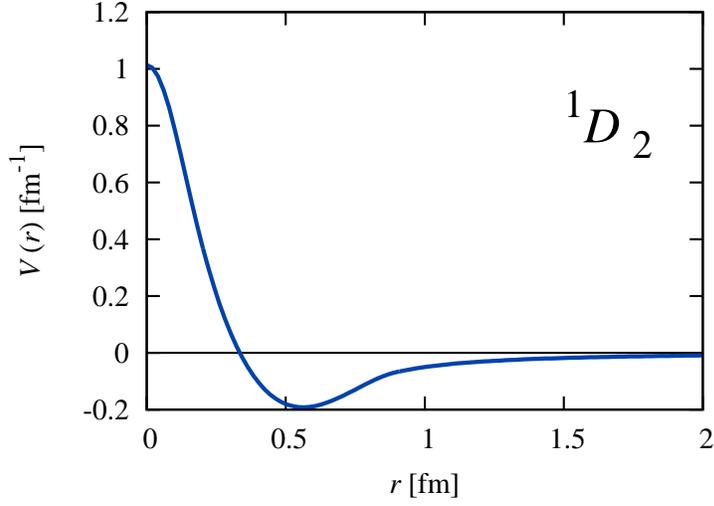}}
\end{center}
\caption{Plot of the model potential in the $^{1}D_{2}$ channel. In this
channel $S=0$, $T=1$, $C_{\pi}=-3$, $C_{A}=-60.5$, $C_{B}=-30.0$, $C_{\Pi}=0$,
$C_{D}=0$, $C_{F}=0$, $C_{G}=1.01$, $m_{A}=9m_{\pi}$, $m_{B}=6m_{\pi}$,
$m_{G}=7.02 m_{\pi}$, $C_{1}=1.463$, $C_{2}=-7.332$, $C_{3}=10.384$ and
$C_{4}=-4.585$. We use $R_{\text{Gauss}}=0.2$ fm and $R_{\text{Exch.}}=0.9$
fm.}%
\label{fig:OPE1D2}%
\end{figure}

\begin{figure}[ptb]
\begin{center}
\resizebox{100mm}{!}{\includegraphics{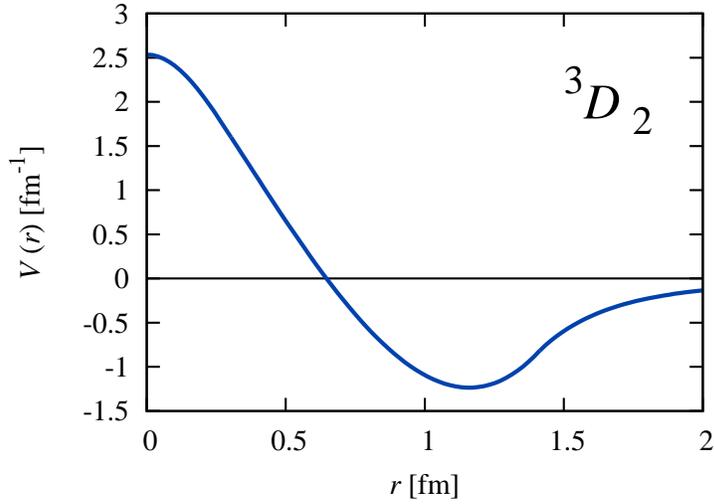}}
\end{center}
\caption{Plot of the model potential in the $^{3}D_{2}$ channel. In this
channel $S=1$, $T=0$, $C_{\pi}=-3$, $C_{A}=-660$, $C_{B}=-1140$, $C_{\Pi}=-3$,
$C_{D}=-930$, $C_{F}=-927$, $C_{G}=2.53$, $m_{A}=m_{D}=8m_{\pi}$, $m_{B}%
=m_{F}=5m_{\pi}$, $m_{G}=3.25 m_{\pi}$, $C_{1}=2.979$, $C_{2}=-4.026$,
$C_{3}=-2.455$ and $C_{4}=2.409$. \ We use $R_{\text{Gauss}}=0.3$ fm and
$R_{\text{Exch.}}=1.4$ fm.}%
\label{fig:OPE3D2}%
\end{figure}

The short-range part is a Gaussian function
\begin{equation}
V_{\text{Gauss}}(r)=C_{G}e^{-m_{G}^{2} r^{2}}. \label{PotentialModelGaussian}%
\end{equation}
The intermediate-range part of the potential is a cubic spline use to connect
the short- and long-range regions,%
\begin{equation}
V_{\text{Spline}}(r)=C_{1}+C_{2}r+C_{3}r^{2}+C_{4}r^{3}.
\label{PotentialModelIntermediate}%
\end{equation}
The long-range part consists of the usual one-pion exchange potential together
with two additional heavy meson exchange terms,%
\begin{equation}
V_{\text{Exch.}}(r)=V_{C}^{\pi,A,B}(r)+S_{12}V_{T}^{\pi,D,F}(r).
\label{PotentialModelOPE}%
\end{equation}
The central part of the potential is composed of Yukawa functions
\begin{equation}
V_{C}^{\pi,A,B}(r)=\frac{g_{\pi N}^{2}}{12\pi}\left(  \frac{m_{\pi}}{2M_{N}%
}\right)  ^{2}\left\{  C_{\pi}\frac{e^{-m_{\pi}r}}{r}+C_{A}\frac{e^{-m_{A}r}%
}{r}+C_{B}\frac{e^{-m_{B}r}}{r}\right\}  , \label{PotentialModel}%
\end{equation}
and the tensor part of the potential has the form%
\begin{align}
V_{T}^{\pi,D,F}(r)  &  =\frac{g_{\pi N}^{2}}{12\pi}\left(  \frac{m_{\pi}%
}{2M_{N}}\right)  ^{2}\left\{  C_{\Pi}\left[  1+\frac{3}{m_{\pi}r}+\frac
{3}{(m_{\pi}r)^{2}}\right]  \frac{e^{-m_{\pi}r}}{r}\right.  \hspace
{5.5cm}\nonumber\\
&  \left.  +C_{D}\left[  1+\frac{3}{m_{D}r}+\frac{3}{(m_{D}r)^{2}}\right]
\frac{e^{-m_{D}r}}{r}+C_{F}\left[  1+\frac{3}{m_{F}r}+\frac{3}{(m_{F}r)^{2}%
}\right]  \frac{e^{-m_{F}r}}{r}\right\}  .
\end{align}
Here $C_{\pi}=(2\vec{S}^{2}-3)(2\vec{T}^{2}-3)$, $C_{\Pi}=2\vec{T}^{2}-3$,
$g_{\pi N}=13.0$, $m_{\pi}=140\,\text{MeV}$, and $M_{N}=938.0$ MeV. The
coefficients not part of the one-pion exchange potential are used as free
parameters to reproduce the physical low-energy scattering parameters. \ Due
to the abundance of free parameters, the fit process is not unique. \ But in
each case we attempt to qualitatively reproduce the shape of the NijmegenII
potentials \cite{Stoks1993NijII}. \ And in each case the heavy meson masses
are kept significantly larger than the pion mass. \ In Fig.~(\ref{fig:OPE1P1})
we show the potential in the $^{1}P_{1}$ channel. \ \ Fig.~(\ref{fig:OPE1D2})
shows the potential in the $^{1}D_{2}$ channel, and Fig.~(\ref{fig:OPE3D2})
shows the potential in the $^{3}D_{2}$ channel.



After having recovered the physical low-energy scattering parameters, we now
multiply an additional step function to the potential,%
\begin{equation}
V(r)\rightarrow V(r)\theta(R-r),
\end{equation}
which removes the tail of the potential beyond range $R$. \ We then
recalculate the low-energy scattering parameters with this modification. \ The
results are shown in Table \ref{table:CausalityBounds} for the $^{1}P_{1}$,
$^{1}D_{2}$, and $^{3}D_{2}$ channels. \ The causal ranges for the $^{1}D_{2}$
and $^{3}D_{2}$ channels are quite large for the physical scattering data,
$4.0$ and $4.9$ fm respectively. \ But if we remove the tail of the model
potential at $R=5$ fm, the causal ranges drop to $2.4$ and $2.7$ fm
respectively. \begin{table}[tbh]
\caption{The potential range dependence of the causal range in various
channels.}%
\label{table:CausalityBounds}
\centering%
\begin{tabular}
[c]{||c|c|c|c|c|c||}\hline\hline
\backslashbox{Causal range}{Potential range} & $R=2\ \text{fm}$ &
$R=5\ \text{fm}$ & $R=12\ \text{fm}$ & $R=15\ \text{fm}$ & $R=50\ \text{fm}%
$\\\hline\hline
$R_{^{1}P_{1}}^{b}$ & 0.4 & 0.4 & 0.3 & 0.3 & 0.3\\\hline
$R_{^{1}D_{2}}^{b}$ & 2.0 & 2.4 & 3.8 & 4.0 & 4.0\\\hline
$R_{^{3}D_{2}}^{b}$ & 1.3 & 2.7 & 4.7 & 4.9 & 4.9\\\hline\hline
\end{tabular}
\end{table}

The tail of the model potential is dominated by the one-pion exchange
potential. \ Even though the numerical size of the one-pion exchange potential
is small at distances of $5$ fm, these numerical results show clearly that the
one-pion exchange tail is controlling the size of the causal range. \ The
one-pion exchange potential tail appears to be the source of the large values
for $R^{b}$ in higher partial waves where the central one-pion exchange tail
is attractive.

\section{Summary and Conclusions}

In this paper we have derived the constraints of causality and unitarity for
neutron-proton scattering for all spin channels up to $J=3$. \ We have defined
and calculated interaction length scales which we call the causal range,
$R^{b}$, and the Cauchy-Schwarz range, $R^{\text{C-S}}$. \ The causal range is
the minimum value for $r$ such that the causal bounds,%
\begin{equation}
f_{J-1}(r)=b_{J-1}(r)-2q_{0}^{2}\frac{\Gamma(J+\frac{1}{2})\Gamma(J+\frac
{3}{2})}{\pi}\left(  \frac{r}{2}\right)  ^{-2J-1}-r_{J-1}\geq0,
\end{equation}%
\begin{equation}
g_{J+1}(r)=b_{J+1}(r)+\frac{2q_{0}^{2}}{a_{J+1}^{2}}\frac{\pi}{\Gamma
(J+\frac{1}{2})\Gamma(J+\frac{3}{2})}\left(  \frac{r}{2}\right)
^{2J+1}-r_{J+1}\geq0,
\end{equation}
are satisfied. \ For uncoupled channels these bounds simplify to the form%
\begin{equation}
f_{L}(r)=g_{L}(r)=b_{L}(r)\geq0.
\end{equation}
For coupled channels the Cauchy-Schwarz range is the minimum value for $r$
satisfying the causal bounds as well as the Cauchy-Schwarz inequality,%
\begin{equation}
f_{J-1}(r)g_{J+1}(r)\geq\left[  h_{J}(r)\right]  ^{2}.
\end{equation}
If one reproduces the physical scattering data using strictly finite range
interactions, then the range of these interactions must be larger than $R^{b}$
and $R^{\text{C-S}}$. \ From these bounds we have derived the general result
that non-vanishing partial-wave mixing cannot be reproduced with zero-range
interactions. \ As the range of the interaction goes to the zero, the
effective range for the lower partial-wave channel is driven to negative infinity.

This finding has consequences for pionless effective theory where the range of
the interactions is set entirely by the value of the cutoff momentum. \ If the
cutoff momentum is too high, then it is impossible to obtain the correct
threshold physics in coupled channels without violating causality or
unitarity. \ In some channels we find that the causal range and Cauchy-Schwarz
range are as large $5$ fm. \ We have shown that these large values are driven
by the tail of the one-pion exchange potential. \ In these channels the
problems will be even more severe, and the cutoff momentum will need to be
rather low in order to reproduce the physical scattering data in pionless
effective field theory. \ How low this cutoff momentum must be depends on the
particular regularization scheme.

We should note that all of these mixing observables are non-vanishing only
when one reaches higher orders in the power counting expansion, and there is
no direct impact on pionless effective field theory calculations at lower
orders. \ See, for example, Ref.~\cite{Chen:1999tn} for details on power
counting in pionless effective field theory. \ In the zero-range limit, the
term which drives the negative divergence of the effective range parameter
$r_{J-1}$ is%
\begin{equation}
-2q_{0}^{2}\frac{\Gamma(J+\frac{1}{2})\Gamma(J+\frac{3}{2})}{\pi}\left(
\frac{r}{2}\right)  ^{-2J-1}. \label{q_0^2}%
\end{equation}
At leading order there is no divergence since there is no partial wave mixing
and $q_{0}=0$. \ If higher-order terms are iterated non-perturbatively as in
Ref.~\cite{vanKolck:1998bw}, then the divergence appears at order $Q^{2}$, the
first order at which $q_{0}$ is non-vanishing. \ If higher-order terms are
iterated order-by-order in perturbation theory, then the term in
Eq.~(\ref{q_0^2}) appears at order $Q^{4}$. \ This is one order higher than
the analysis presented in Ref.~\cite{Chen:1999tn}, and we predict that
zero-range divergences in $r_{J-1}$ will first appear at this order.

It important to note that if one works order-by-order in perturbation
theory, then the constraints of causality and unitarity always appear
somewhat hidden.  At every order in the effective field theory
calculation there are new operator coefficients which appear and
are determined by matching to physical data.  There are no
obstructions to setting these operator coefficients to reproduce
physical values.

It is only when one iterates the new interactions, i.e., by solving
the Schr{\"o}dinger equation, that non-linear dependencies on
the operator coefficients appear.  In this case one finds that the
constraints of causality and unitarity give necessary conditions
for keeping the operator coefficients real.  Once we fix the
regularization, the bound corresponds with branch cuts of the
effective theory when viewed as a function of physical scattering
parameters.

These branch cuts cannot be seen at any finite order in
perturbation theory.  However a nearby branch point may
spoil the convergence of the perturbative expansion.  In
this context, our causality and unitarity bounds can be
viewed as setting physical constraints for the convergence of
perturbative calculations in pionless effective field theory.

If the cutoff is taken too high, a branch cut develops which
jeopardizes the convergence of the perturbative calculation.
Similarly if one does calculations using dimensional
regularization, then the renormalization scale sets the scale
at which the infrared and ultraviolet physics are regulated
\cite{Georgi:1992xg}.  Similar problems with perturbative
convergence would arise if the renormalization scale is
taken too high.

There is much theoretical interest in the connection between dilute neutron
matter and the universal physics of fermions in the unitarity limit
\cite{Gezerlis:2007fs,Borasoy:2007vk,Epelbaum:2008vj,Lee:2008xs,Gezerlis:2009iw,Wlazlowski:2009yi,Forbes:2012a}%
. \ In the limit of isospin symmetry our analysis of the isospin triplet
channels can be applied to neutron-neutron scattering in dilute neutron
matter. \ In this paper we have shown there are intrinsic length scales
associated with the causal range and the Cauchy-Schwarz range. \ When the
average separation between neutrons is smaller than these length scales, one
expects non-universal behavior controlled by the details of the
neutron-neutron interactions. For the $^{1}S_{0}$ channel, $R^{b}=1.3$~fm.
\ For the $^{3}P_{2}$ channel, $R^{b}=2.2$~fm, and for the $^{3}F_{2}$
channel, $R^{b}=1.7$~fm. \ For $^{3}P_{2}$-$^{3}F_{2}$ mixing, we find
$R^{\text{C-S}}=4.7$~fm. \ We see that the physics of $^{3}P_{2}$-$^{3}F_{2}$
mixing will become non-universal at lower densities than the $^{1}S_{0}$
interactions. \ In particular the densities where $^{3}P_{2}$ superfluidity is
expected to occur will be well beyond this universal regime.

\section*{Acknowledgements}

We thank Hans-Werner Hammer, Sebastian K\"{o}nig, Bira van Kolck,  Evgeny Epelbaum, Jambul Gegelia, and Hermann Krebs for
useful discussions. \ Financial support from U.S. Department of Energy
(DE-FG02-03ER41260) is acknowledged. S.~E. is supported by a Turkish
Government Ministry of National Education Fellowship.

\appendix

\section{Functions}

\label{append:A}

\subsection{The Riccati-Bessel Functions}

\label{append:RiccatiBessel}

$S_{J\pm1}(r)$ and $C_{J\pm1}(r)$ are Riccati-Bessel functions of the first
and second kind, which are defined in terms of the Bessel
functions as%
\begin{align}
S_{J\pm1}(r) =\sqrt{\frac{\pi r}{2}}J_{J\pm1+\frac{1}{2}%
}(r) =\sqrt{\pi}r^{J\pm1+1}\sum_{n=0}^{\infty}\frac{i^{2n}2^{-2n-J\mp1-1}%
}{\Gamma(n+1)\Gamma(n+J\pm1+\frac{3}{2})}r^{2n}, \label{eqn:firstricbes}%
\end{align}%
\begin{align}
C_{J\pm1}(r)  & =-\sqrt{\frac{\pi r}{2}}Y_{J\pm1+\frac{1}{2}%
}(r)\nonumber\\
&  =\frac{r^{-J\mp1}}{\sqrt{\pi}}\Gamma(-J\mp1+\frac{1}{2})\Gamma(J\pm
1+\frac{1}{2})\sum_{n=0}^{\infty}\frac{i^{2n}2^{-2n+J\pm1}}{\Gamma
(n+1)\Gamma(n-J\mp1+1/2)}r^{2n}. \label{eqn:secondricbes}%
\end{align}

The $s(r)$ and $c(r)$ functions in Eq.~(\ref{eqn:expfuncS2}) and
Eq.~(\ref{eqn:expfuncC2}) are written from Eq.~(\ref{eqn:firstricbes}) and
Eq.~(\ref{eqn:secondricbes}),%
\begin{align}
s_{0,J\pm1}(r)  &  =\frac{\sqrt{\pi}}{\Gamma\left(  J\pm1+\frac{3}{2}\right)
}\left(  \frac{r}{2}\right)  ^{J\pm1+1},\label{eqn:expfuncS1term}\\
s_{2,J\pm1}(r)  &  =-\frac{\sqrt{\pi}}{\Gamma\left(  J\pm1+\frac{5}{2}\right)
}\left(  \frac{r}{2}\right)  ^{J\pm1+3},\label{eqn:expfuncS2term}\\
c_{0,J\pm1}(r)  &  =\frac{\Gamma\left(  J\pm1+\frac{1}{2}\right)  }{\sqrt{\pi
}}\left(  \frac{r}{2}\right)  ^{-J\mp1},\label{eqn:expfuncC1term}\\
c_{2,J\pm1}(r)  &  =\frac{\Gamma\left(  J\pm1-\frac{1}{2}\right)  }{\sqrt{\pi
}}\left(  \frac{r}{2}\right)  ^{-J\mp1+2}. \label{eqn:expfuncC2term}%
\end{align}

\subsection{Wronskians of Wave Functions}

\label{append:Wronskiansof functions}

Here we calculate Wronskians of the $U(r)$ and $V(r)$ wave functions, and
Wronskians of all possible combination of the $s_{0}(r)$, $s_{2}(r)$,
$c_{0}(r)$ and $c_{2}(r)$ functions. \ Wronskians of $U_{\alpha}(r)$ and
$V_{\alpha}(r)$ for the non-interacting region $r\geq R$ are%
\begin{align}
W[U_{a\alpha}(r),U_{b\alpha}(r)]  &  =(k_{a}^{2}-k_{b}^{2})\Big\{\frac{1}%
{2}r_{J-1}W[s_{0}(r),c_{0}(r)]_{J-1}\nonumber\\
&  +\frac{1}{a_{J-1}^{2}}W[s_{2}(r),s_{0}(r)]_{J-1}+\frac{1}{a_{J-1}}%
W[c_{0}(r),s_{2}(r)]_{J-1}\nonumber\\
&  +\frac{1}{a_{J-1}}W[s_{0}(r),c_{2}(r)]_{J-1}+W[c_{2}(r),c_{0}%
(r)]_{J-1}\Big\}\nonumber\\
&  +\mathcal{O}(k_{a}^{4})+\mathcal{O}(k_{b}^{4}), \label{eqn:wronskUalp}%
\end{align}%
\begin{equation}
W[V_{a\alpha}(r),V_{b\alpha}(r)]=(k_{a}^{2}-k_{b}^{2})q^{2}W[c_{2}%
(r),c_{0}(r)]_{J+1}+\mathcal{O}(k_{a}^{4})+\mathcal{O}(k_{b}^{4}).
\label{eqn:wronskValp}%
\end{equation}
Wronskians of the $\beta$-state wave functions are
\begin{equation}
W[U_{a\beta}(r),U_{b\beta}(r)]=(k_{a}^{2}-k_{b}^{2})q^{2}\frac{1}{a_{J+1}^{2}%
}W[s_{2}(r),s_{0}(r)]_{J-1}+\mathcal{O}(k_{a}^{4})+\mathcal{O}(k_{b}^{4}),
\label{eqn:wronskUbet}%
\end{equation}%
\begin{align}
W[V_{a\beta}(r),V_{b\beta}(r)]  &  =(k_{a}^{2}-k_{b}^{2})\Big\{\frac{1}%
{2}r_{J+1}W[s_{0}(r),c_{0}(r)]_{J+1}\nonumber\\
&  +\frac{1}{a_{J+1}^{2}}W[s_{2}(r),s_{0}(r)]_{J+1}+\frac{1}{a_{J+1}}%
W[c_{0}(r),s_{2}(r)]_{J+1}\nonumber\\
&  +\frac{1}{a_{J+1}}W[s_{0}(r),c_{2}(r)]_{J+1}+W[c_{2}(r),c_{0}%
(r)]_{J+1}\Big\}\nonumber\\
&  +\mathcal{O}(k_{a}^{4})+\mathcal{O}(k_{b}^{4}). \label{eqn:wronskVbet}%
\end{align}
Wronskian of the combinations of the $\alpha$ and $\beta$-states are
\begin{align}
W[U_{a\alpha}(r),U_{b\beta}(r)]  &  =q\frac{1}{a_{J+1}}W[c_{0}(r),s_{0}%
(r)]_{J-1}-k_{a}^{2}\Big\{q\frac{1}{a_{J-1}a_{J+1}}W[s_{2}(r),s_{0}%
(r)]_{J-1}\nonumber\\
&  -q\frac{1}{a_{J+1}}W[c_{2}(r),s_{0}(r)]_{J-1}\Big\}+k_{b}^{2}%
\Big\{q\frac{1}{a_{J-1}a_{J+1}}W[s_{2}(r),s_{0}(r)]_{J-1}\nonumber\\
&  -q\frac{1}{a_{J+1}}W[s_{2}(r),c_{0}(r)]_{J-1}+q\frac{r_{J+1}}{2}%
W[s_{0}(r),c_{0}(r)]_{J-1}\nonumber\\
&  -q_{1}\frac{1}{a_{J+1}}W[s_{0}(r),c_{0}(r)]_{J-1}\Big\}+\mathcal{O}%
(k^{4}),\hspace{4.5cm} \label{eqn:wronskUaalpUbbet}%
\end{align}%
\begin{align}
W[U_{a\beta}(r),U_{b\alpha}(r)]  &  =-q\frac{1}{a_{J+1}}W[c_{0}(r),s_{0}%
(r)]_{J-1}+k_{b}^{2}\Big\{q\frac{1}{a_{J-1}a_{J+1}}W[s_{2}(r),s_{0}%
(r)]_{J-1}\nonumber\\
&  -q\frac{1}{a_{J+1}}W[c_{2}(r),s_{0}(r)]_{J-1}\Big\}-k_{a}^{2}%
\Big\{q\frac{1}{a_{J-1}a_{J+1}}W[s_{2}(r),s_{0}(r)]_{J-1}\nonumber\\
&  -q\frac{1}{a_{J+1}}W[s_{2}(r),c_{0}(r)]_{J-1}+q\frac{r_{J+1}}{2}%
W[s_{0}(r),c_{0}(r)]_{J-1}\nonumber\\
&  -q_{1}\frac{1}{a_{J+1}}W[s_{0}(r),c_{0}(r)]_{J-1}\Big\}+\mathcal{O}%
(k^{4}),\hspace{4.5cm} \label{eqn:wronskUabetUbalp}%
\end{align}%
\begin{align}
W[V_{a\alpha}(r),V_{b\beta}(r)]  &  =-q\frac{1}{a_{J+1}}W[c_{0}(r),s_{0}%
(r)]_{J+1}-{k_{a}^{2}}\Big\{q_{1}\frac{1}{a_{J+1}}W[c_{0}(r),s_{0}%
(r)]_{J+1}\nonumber\\
&  +q\frac{1}{a_{J+1}}W[c_{2}(r),s_{0}(r)]_{J+1}-qW[c_{2}(r),c_{0}%
(r)]_{J+1}\Big\}\nonumber\\
&  +{k_{b}^{2}}\Big\{q\frac{r_{J+1}}{2}W[c_{0}(r),s_{0}(r)]_{J+1}-q\frac
{1}{a_{J+1}}W[c_{0}(r),s_{2}(r)]_{J+1}\nonumber\\
&  +qW[c_{0}(r),c_{2}(r)]_{J+1}\Big\}+\mathcal{O}(k^{4}),
\label{eqn:wronskVaalpVbbet}%
\end{align}%
\begin{align}
W[V_{a\beta}(r),V_{b\alpha}(r)]  &  =q\frac{1}{a_{J+1}}W[c_{0}(r),s_{0}%
(r)]_{J+1}+{k_{b}^{2}}\Big\{q_{1}\frac{1}{a_{J+1}}W[c_{0}(r),s_{0}%
(r)]_{J+1}\nonumber\\
&  +q\frac{1}{a_{J+1}}W[c_{2}(r),s_{0}(r)]_{J+1}-qW[c_{2}(r),c_{0}%
(r)]_{J+1}\Big\}\nonumber\\
&  -{k_{a}^{2}}\Big\{q\frac{r_{J+1}}{2}W[c_{0}(r),s_{0}(r)]_{J+1}-q\frac
{1}{a_{J+1}}W[c_{0}(r),s_{2}(r)]_{J+1}\nonumber\\
&  +qW[c_{0}(r),c_{2}(r)]_{J+1}\Big\}+\mathcal{O}(k^{4}).
\label{eqn:wronskVabetVbalp}%
\end{align}

We now calculate Wronskians of the $s_{0}(r)$, $s_{2}(r)$, $c_{0}(r)$ and
$c_{2}(r)$ functions. \ We find
\begin{equation}
W[s_{0}(r),s_{2}(r)]_{J}=-\frac{\pi}{\Gamma\left(  \frac{3}{2}+J\right)
\Gamma\left(  \frac{5}{2}+J\right)  }\left(  \frac{r}{2}\right)  ^{3+2J},
\label{eqn:wronskS0S2}%
\end{equation}%
\begin{equation}
W[s_{0}(r),c_{0}(r)]_{J}=-1, \label{eqn:wronskS0C0}%
\end{equation}%
\begin{equation}
W[s_{0}(r),c_{2}(r)]_{J}=-\frac{r^{2}}{2+4J}, \label{eqn:wronskS0C2}%
\end{equation}%
\begin{equation}
W[s_{2}(r),c_{0}(r)]_{J}=\frac{r^{2}}{2+4J}, \label{eqn:wronsks2c0}%
\end{equation}%
\begin{equation}
W[s_{2}(r),c_{2}(r)]_{J}=\frac{r^{4}}{16J(J+1)-12}, \label{eqn:wronsks2c2}%
\end{equation}%
\begin{equation}
W[c_{0}(r),c_{2}(r)]_{J}=-\frac{\Gamma\left(  -\frac{1}{2}+J\right)
\Gamma\left(  \frac{1}{2}+J\right)  }{\pi}\left(  \frac{r}{2}\right)  ^{1-2J}.
\label{eqn:wronskc0c2}%
\end{equation}
It should be noted that $W[f(r),g(r)]=-W[f(r),g(r)]$.

\section{Relations between Eigenphase and Nuclear Bar Parameterizations}

\label{append:B}The scattering matrix in terms of the eigenphase parameters
was given in Eq.~(\ref{eqn:scattmat}). \ The scattering matrix in terms of the
nuclear bar parameters is
\begin{equation}
S=\left(
\begin{array}
[c]{cc}%
e^{2i\bar{\delta}_{\alpha}}\cos2\bar{\varepsilon} & ie^{i\left(  \bar{\delta
}_{\alpha}+\bar{\delta}_{\beta}\right)  }\sin2\bar{\varepsilon}\\
ie^{i\left(  \bar{\delta}_{\alpha}+\bar{\delta}_{\beta}\right)  }\sin
2\bar{\varepsilon} & e^{2i\bar{\delta}_{\beta}}\cos2\bar{\varepsilon}%
\end{array}
\right)  .
\end{equation}
Here $\overline{\delta}_{\alpha}$, $\overline{\delta}_{\beta}$ and
$\overline{\varepsilon}$ are the nuclear bar phase shifts and mixing angle
\cite{Stapp1956}. \ The relations between the eigenphase and the nuclear bar
parameters are
\begin{align}
\sin(\delta_{\alpha}-\delta_{\beta})  &  =\frac{\sin2\overline{\varepsilon}%
}{\sin2{\varepsilon}},\\
\delta_{\alpha}+\delta_{\beta}  &  =\overline{\delta}_{\alpha}+\overline
{\delta}_{\beta},\\
\tan2\varepsilon &  =\frac{\tan2\overline{\varepsilon}}{\sin(\overline{\delta
}_{\alpha}-\overline{\delta}_{\beta})}. \label{eqn:eigen_nuclearbar}%
\end{align}

The two-channel effective range expansion is defined slightly differently in
the eigenphase and the nuclear bar parameterizations. \ In the eigenphase
parameterization,
\begin{equation}
k^{L_{ij}+\frac{1}{2}}U\hat{\mathbf{K}}^{-1}U^{-1}k^{L_{ij}+\frac{1}{2}%
}=-\frac{1}{a_{ij}}+\frac{1}{2}r_{ij}k^{2}+\mathcal{O}(k^{4}),
\label{eqn:EFE_BB}%
\end{equation}
and in the nuclear bar parameterization,%
\begin{equation}
k^{L_{ij}+\frac{1}{2}}\hat{\mathbf{K}}^{-1}k^{L_{ij}+\frac{1}{2}}=-\frac
{1}{\bar{a}_{ij}}+\frac{1}{2}\bar{r}_{ij}k^{2}+\mathcal{O}(k^{4}).
\label{eqn:EFE_SYM}%
\end{equation}
Therefore, by straightforward calculations we find the following relations
among the threshold scattering parameters,%
\begin{align}
a_{\alpha}=  &  \bar{a}_{\alpha},\label{eqn:ScattLenAlpha}\\
r_{\alpha}=  &  \bar{r}_{\alpha}+\frac{2\bar{q}_{0}\bar{q}_{1}}{\bar
{a}_{\alpha}}+\frac{\bar{q}^{2}\bar{r}_{\beta}}{\bar{a}_{\alpha}^{2}%
},\label{eqn:EffecRangAlpha}\\
a_{\beta}=  &  \bar{a}_{\beta}-\frac{\bar{q}_{0}^{2}}{\bar{a}_{\alpha}%
},\label{eqn:ScattLenBeta}\\
r_{\beta}=  &  \bar{r}_{\beta},\label{eqn:EffectRangBeta}\\
q_{0}=  &  \frac{\bar{q}_{0}}{\bar{a}_{\alpha}},\label{eqn:MixAngFirst}\\
q_{1}=  &  \frac{\left(  \bar{a}_{\beta}\bar{a}_{\alpha}-\bar{q}_{0}%
^{2}\right)  \left(  \bar{a}_{\alpha}\bar{q}_{1}+\bar{r}_{\beta}\bar{q}%
_{0}\right)  }{2\bar{a}_{\alpha}^{2}}. \label{eqn:MixAngSecond}%
\end{align}
For the uncoupled channels $q_{0}$ and $q_{1}$ are zero, and these relations
become $a_{\alpha}=\bar{a}_{\alpha}$, $r_{\alpha}=\bar{r}_{\alpha}$,
$a_{\beta}=\bar{a}_{\beta}$, and $r_{\beta}=\bar{r}_{\beta}$.

\newpage

\bibliographystyle{apsrev}
\bibliography{Reference}

\end{document}